\RequirePackage{fix-cm}
\documentclass[twocolumn]{svjour3}          

\smartqed  

\usepackage{graphicx}
\usepackage{url}

\usepackage{hyperref}
\hypersetup{breaklinks}
\usepackage{breakurl}
\usepackage{latexsym}
\usepackage{newtxmath}
\usepackage{amsmath}
\usepackage{tabulary, booktabs}
\usepackage{colortbl}
\usepackage{multirow}
\usepackage[numbers]{natbib}

\newcolumntype{S}{>{\arraybackslash}m{0.55cm}}
\newcolumntype{M}{>{\arraybackslash}m{1.3cm}}
\newcolumntype{W}{>{\centering\arraybackslash}m{3.5cm}}
\newcolumntype{Q}{>{\centering\arraybackslash}m{1.6cm}}
\newcolumntype{L}{>{\centering\arraybackslash}m{2cm}}
\newcolumntype{B}{>{\centering\arraybackslash}m{5cm}}
\newcolumntype{G}{>{\arraybackslash}m{8cm}}

\journalname{ArXiv}
\begin{document}

\title{Assessing Holistic Impacts of Major Events on the Bitcoin Blockchain Network \thanks{This research was supported in part by the National Science Foundation under grant CNS 1461133.}
}


\author{Anthony Luo          \and
        Dianxiang Xu 
}

\institute{Anthony Luo \at
              Fu Foundation School of Engineering and Applied Science\at
              Columbia University \at
              New York, NY 10027, USA \at
              \email{anthony.luo@columbia.edu}           
           \and
           Dianxiang Xu \at
              Department of Computer Science and Electrical Engineering \at
              University of Missouri - Kansas City \at
            Kansas City, MO 64110, USA \at \email{dxu@umkc.edu} 
}

\date{}

\maketitle

\begin{abstract}
As the pioneer of blockchain technology, Bitcoin is the most popular cryptocurrency to date. Given its dramatic price spikes (and crashes) along with the never-ending news from SEC regulations to security breaches, there seems to be a lack of understanding about the dynamics of cryptocurrencies. These dynamics are believed to be affected by various political, security, financial, and regulatory events. In this paper, we present an efficient framework for holistic analysis of cryptocurrency fluctuations by introducing the Impact-Score metric to distinguish event-induced changes from normal variations. We have applied our framework to 16 major worldwide events and the Bitcoin blockchain network (defined as Bitcoin transaction and users, blockchain data, and memory pool data) from 2016-2018. The results show that a majority of the events are correlated with substantial network changes. We observed roughly generalizable correlations between event types (e.g. financial events) and sub-structures of the Bitcoin blockchain network. Subgroups of these events have strongly consistent temporal impacts on specific facets (e.g. activity or fees) of the Bitcoin ecosystem. Furthermore, we demonstrate the robustness of our process by correlating a majority of spikes in network/subnetwork change with major events.
\keywords{Bitcoin \and blockchain \and cryptocurrency \and data analytics}
\end{abstract}

\section{Introduction}
\label{sec:intro}
Blockchain is an enabling technology that has the potential to revolutionize industries from finance to healthcare to energy. Cryptocurrencies, a type of blockchain technology, are worth over 200 billion USD today and over half a trillion USD at their peak. Hundreds of corporations \textemdash startups and veterans alike \textemdash have begun to explore applications of blockchain technology. However, there is a lack of clear understanding about the dynamics of these cryptocurrencies. They are surrounded and affected by events from security breaches, forks, speculation, news, regulation, and unrelated world events. For example, the price of Bitcoin increased 3\% following the election of Donald Trump. A quantitative measure of holistic event impact would improve our understanding about the dynamics of cryptocurrencies. To this end, there are two major challenges. First, how can we quantitatively capture the specific states of a cryptocurrency over a given time period from the holistic perspective of users, transactions, blockchain, and network? Second, how can we quantitatively measure aspect-specific changes caused by major events beyond normal fluctuations?

This paper focuses on Bitcoin \cite{nakamoto2008bitcoin}, the first and most popular cryptocurrency. We investigate the dynamics of Bitcoin with regards to major worldwide events. We define the Bitcoin blockchain network as the aggregate of Bitcoin transaction and users, blockchain data, and memory pool (also called mempool), and subnetworks with respect to the Bitcoin activity, fee, and transaction. The ideal approach to quantitatively measure changes, given unlimited resources and time, would be to utilize graph matching or graphlet comparisons between graph representations of network states where a node represents a Bitcoin address and an edge represents a transaction. However, this approach is impractical as graph matching, graphlet comparison, and network parameter calculations are extremely computationally intensive due to large graph sizes. 

We propose an efficient framework for event-induced change analysis. To address the first challenge, we analyze statistic features of key attributes of Bitcoin to capture the state of the Bitcoin blockchain network or its subnetworks over a given period of time. We introduce Blockchain Network Structure (BNS) Vectors to represent these features to facilitate network state comparison without using graph matching. To address the second challenge, we introduce the Impact-Score metric, which standardizes event-associated BNS vector changes to background BNS vector changes. This allows us to measure event impact by separating normal fluctuations from event-induced changes. We have applied our framework to 16 major security, regulatory, political, and financial events from 2016-2018 on the Bitcoin blockchain network. We observed a substantial change in the overall Bitcoin blockchain network, or its subnetworks at a time associated with 15 of the 16 events. We also observed a rough correlation between different event types and shifts of subnetworks. Further analysis of three events - the election of Trump, the Bitcoin-Bitcoin Cash Hard Fork, and the announcement of a Bitcoin ban in South Korea - reveal specific temporal changes in various subnetworks and major shifts in Bitcoin transaction value distributions consistent with predicted user behavior. Events in each subgroup of Bitcoin valuation milestones, Bitcoin/cryptocurrency security, or major FUD-inducing world events, have similar blockchain changes.

We validate our framework for robustness by analyzing substantial changes observed from 2016-2018 that were not associated with the 16 events that we selected. We can correlate world events to a majority of observed substantial changes outside of our 16 events. Furthermore, retrospective analysis of these observed changes reveals further consistencies among event types and impact. 

The rest of the paper is organized as follows. Section \ref{sec:relatedwork} introduces related work. Section \ref{sec:framework} details our framework. Section \ref{sec:casestudy} explains the setup for our case study, the events selected for analysis, and results obtained. Section \ref{sec:discussion} presents discussion and analysis outside of events selected. Section \ref{sec:conclusion} concludes this paper. 

\section{Related Work}
\label{sec:relatedwork}
This work is an application of data analytics to a novel exploratory study on Bitcoin transactions. 
Existing research has largely focused on topological analysis of the Bitcoin transaction and user networks or the connection between individual indicators of the Bitcoin ecosystem (such as market price) and events or global trends. In contrast, instead of analyzing the Bitcoin blockchain and network from a data first perspective, we present an analysis of the Bitcoin blockchain network by first selecting events for study and then analyzing the Bitcoin blockchain network around these events. Reid and Harrigan \cite{reid2013analysis} were the first to analyze the Bitcoin transaction graph. They focused on user de-anonymization and demonstrated the ability to imply ownership and linkage in a case of Bitcoin theft.  Baumann et al. \cite{baumann2014exploring} conducted an exploratory study on the Bitcoin transaction graph using network analysis and descriptive statistics. They found correlations between exchange rate and user activity and postulated that certain events from 2009-2013 could have explained some shifts in Bitcoin exchange rate. The Bitcoin transaction graph from 2009-2012 was also analyzed through statistical measures by Ron and Shamir \cite{ron2013quantitative}. They also applied clustering methods to the analysis of Bitcoin transaction flows and found certain transaction patterns that are potentially associated with attempts to obfuscate connections between certain addresses. Several others \cite{ober2013structure, meiklejohn2013fistful, androulaki2013evaluating} also analyzed the Bitcoin transaction/user network with regards to anonymity. Spagnuolo et al. \cite{spagnuolo2014bitiodine} created BitIodine, a Bitcoin blockchain analysis framework using user and transaction graphs for forensics analysis. Lischke and Fabian \cite{lischke2016analyzing} analyzed the Bitcoin transaction and economy network aggregated with off-blockchain data such as geolocation through 2013. They presented several findings with regards to geolocation and Bitcoin usage, businesses, gambling, etc. as a whole. Fleder \cite{fleder2015bitcoin} applied external data to the Bitcoin transaction graph in conjunction with clustering to tag addresses and entities. Using PageRank, their graph-analysis framework identified notable activity which they found to be associated with the FBI seizure of Silk Road assets. Kondor et al. \cite{kondor2014inferring} analyzed the topology of the Bitcoin network and found network structural changes associated with the exchange price of Bitcoin. Kondor et. al \cite{kondor2014rich} also investigated the dynamics and topology of the Bitcoin network with regards to wealth distribution across users/entities and Bitcoin flows. Kristoufek \cite{kristoufek2013bitcoin, kristoufek2015main} analyzed the connection between Bitcoin exchange rate and factors such as global events, internet trends, and other phenomena through statistical analysis and wavelet coherence analysis. Feder et al. \cite{feder2018impact}examined the effect of security shocks on Bitcoin exchanges. McGinn et al. \cite{mcginn2016visualizing} developed a large Bitcoin transaction network visualization system which they applied to demonstrate visual characteristics of DDOS attacks and money laundering on the Bitcoin transaction graph. Maesa \cite{maesa2018data} performed topology analysis of the Bitcoin user graph with regards to graph properties such as centrality.

\section{Framework}
\label{sec:framework}
\begin{figure}
\includegraphics[width=0.45\textwidth]{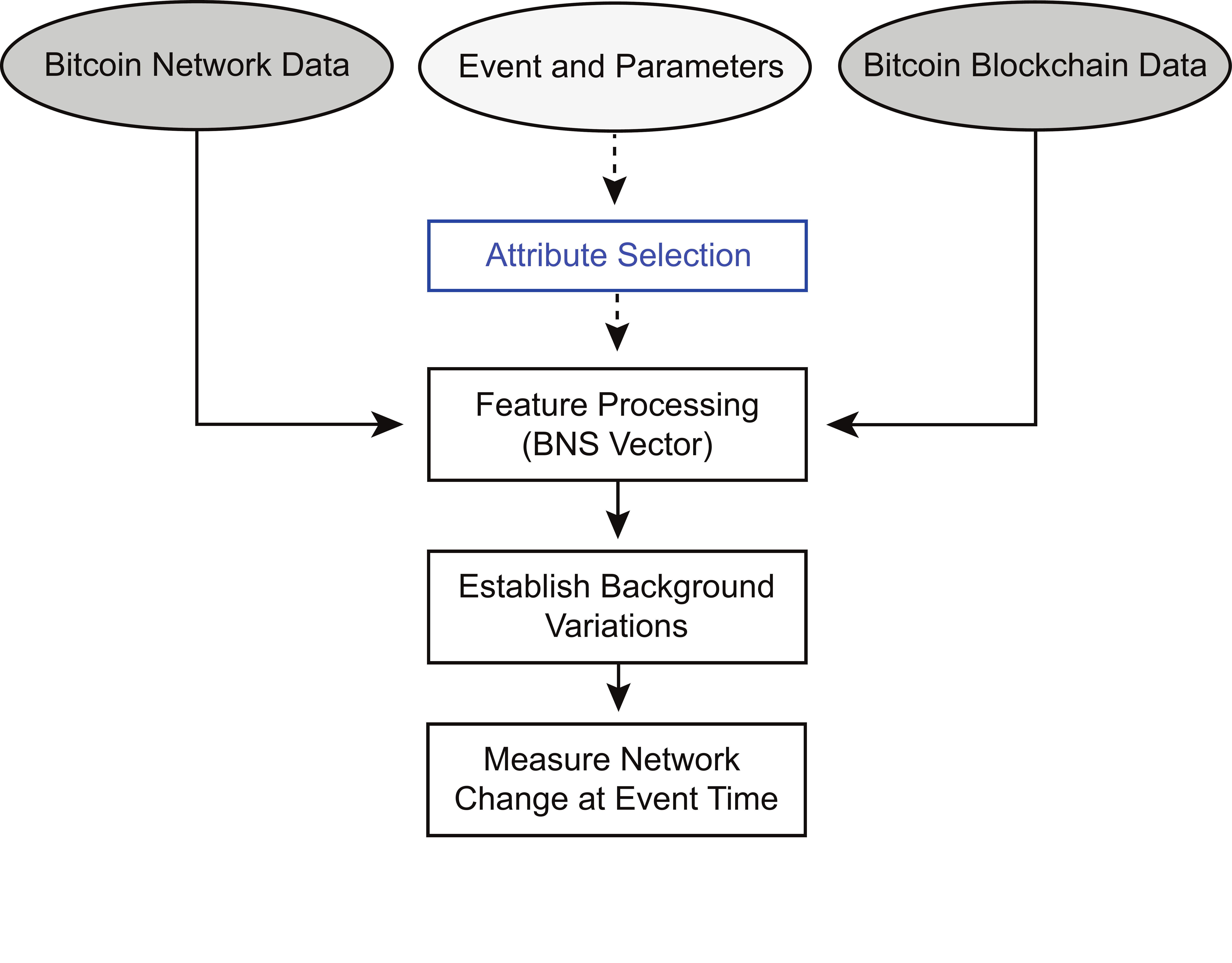}
\caption{Event-Impact Analysis Framework}
\label{fig:framework}   
\end{figure}

Given Bitcoin blockchain data and network data (mempool, blocks, addresses, and transactions), we aim to evaluate the impact of a given event of interest. As shown in Figure \ref{fig:framework}, our framework consists of the following components: attribute selection, feature processing, background fluctuation establishment, and measurement of network change at event time. Sample parameters for a given event include a specific network to be studied, time period, and feature processing configuration. In attribute selection, attributes from Bitcoin blockchain and network data were selected. Each of these attributes is an indicator of some facet of the Bitcoin blockchain network. We use features to capture the distribution of raw data values obtained from each attribute without capturing all the raw data values themselves. We group these features into four feature sets – where each feature set captures the overall Bitcoin blockchain network or one of the three subnetworks we defined (activity, fee, and transaction) respectively. Using the feature set selected, BNS vectors are created to capture the specified network at various points in time. The background or “assumed normal” blockchain network fluctuation two months before and after the event is analyzed and the change in the blockchain network associated with that event is analyzed. The blockchain network change at the event time is then compared to the background fluctuations through the impact score metric to determine event impact significance. Note that we cannot definitively associate events with changes/impact observed in the Bitcoin blockchain network, however we are able to associate events with corresponding impacts with high confidence due to our Impact-Score process (comparing against normal non-event fluctuations). 

\subsection{Attribute Selection}
\label{sec:atrribselect}
The raw Bitcoin blockchain and network data is composed of four general types: transaction, block, address, and mempool. A bitcoin transaction is the transfer of bitcoins from one Bitcoin address or addresses to another Bitcoin address(es). When a transaction is first broadcast, it is added to the Bitcoin mempool, which is a holding pool of transactions that have been broadcast but have not been added to the blockchain. Transactions in the mempool are referred to as “unconfirmed transactions”. The Bitcoin blockchain is a cryptographically connected, chronological chain of 1 MB blocks, where each block contains unique transactions and information regarding the Bitcoin blockchain. Transactions that are in a block are termed “confirmed transactions”. When a new block is mined, transactions contained in that latest block are removed from the mempool (preference is generally given to transactions with higher transaction fees). Table \ref{tab:attributes} lists the 13 attributes selected from these four types of data.

\begin{table*}
\centering
\caption{Attributes of Bitcoin Blockchain and Network Data}
\label{tab:attributes}
\arrayrulecolor{black}
\begin{tabular}{!{\color{black}\vrule}L!{\color{black}\vrule}B!{\color{black}\vrule}G!{\color{black}\vrule}} 
\hline
~                            & Attribute                                                                                                                                                      & Explanation                                                                                                                                                                                                                                                                                                                                                                                \\ 
\hline
\multirow{5}{*}{Transaction} & Transaction Value                                                                                                                                              & The transaction value is the value in Bitcoin (BTC) of a transaction in the Bitcoin blockchain. Bitcoins can be divided down to increments of 1 Satoshi (1 Satoshi = 0.00000001 BTC). For reference, as of August 1, 2018, 1 BTC = \$7,593 USD.                                                                                                                                            \\ 
\cline{2-3}
                             & Transaction Size (bytes)                                                                                                                                       & Transaction size is the size in bytes of a transaction. In most standard transactions, the greater the number of inputs and outputs, the greater the transaction size. Typical transactions are around 250 bytes.                                                                                                                                                                          \\ 
\cline{2-3}
                             & Total Transaction Fees Paid, Fee Rate, Fee Percent                                                       & The transaction fee is the amount (in BTC) attached to a transaction. This is an incentive for miners to add the transaction to the blockchain due to the block size cap of 1 MB. \newline Fees Paid (BTC) = value of fee for the given transaction. \newline Fee Rate (Satoshi/byte) = Fees Paid / Size of Transaction. \newline Fee Percent = Fees Paid / Value of Transaction                                        \\ 
\cline{2-3}
                             & Transactions per Second, Transactions per Block                                                            & Transaction rate is measured in the number of transactions added to the mempool per second and by the number of transactions in a block that is added to the blockchain. Note that during periods of high activity, the transaction rate (transactions being broadcast and added to the mempool) may exceed the number of transactions added to each new block due to the block size cap.  \\ 
\cline{2-3}
                             & Percentage of Non-Standard Transactions                                                                                                                        & Non-Standard transactions are transactions with a non-standard script.                                                                                                                                                                                                                                                                                                                     \\ 
\hline
Block                        & Block Size (MB)                                                                                                                                                & The size of a block in the Bitcoin blockchain in megabytes.                                                                                                                                                                                                                                                                                                                                \\ 
\hline
Addresses                    & Number of unique active addresses over timeframe                                                                                                               & The number of Bitcoin addresses that send or receive a transaction within that time frame. Note that it is common for a single user/entity to have several addresses.                                                                                                                                                                                                                      \\ 
\hline
Memory Pool                  & Mempool Size (bytes), Number of Transactions in the Mempool (Mempool Count), Mempool Growth (bytes/second)  & Due to the block size cap of 1 MB (4 MB after Segwit), the size of the mempool and the number of transactions in the mempool will increase during periods of high activity as there are more transactions being broadcast than can be added to a given block.                                                                                                                               \\
\hline
\end{tabular}
\arrayrulecolor{black}
\end{table*}

\subsection{Feature Processing}
\label{sec:fprocess}
\begin{figure*}
\includegraphics[width=1.05\textwidth]{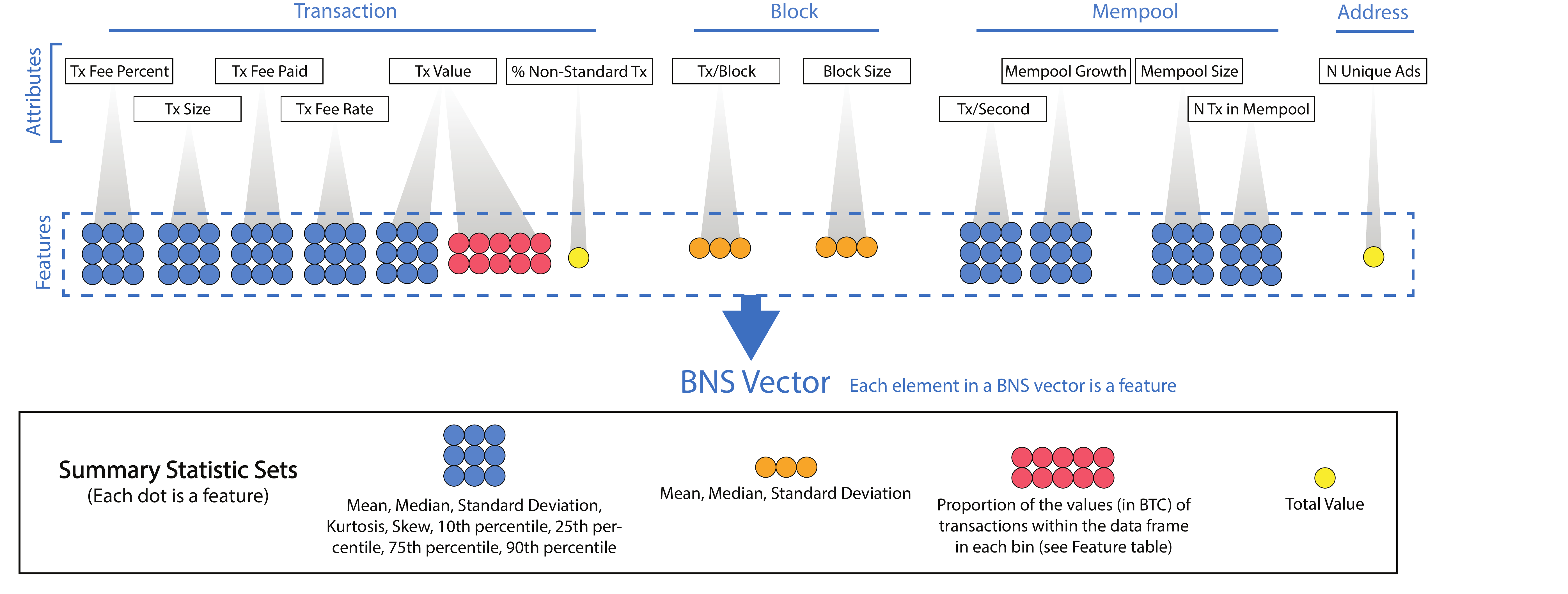}
\caption{Event-Impact Analysis Framework}
\label{fig:aggrprocess}   
\end{figure*}

To reduce the multi-dimensional, size-variant raw data (blockchain and mempool data) over a given time period into a single vector with consistent specifications, we propose summary statistic measures as features to capture the distribution of each attribute according to the data collection unit of the attribute (the frequency at which a value is collected). An attribute with a greater number of raw data values (e.g. data collection unit = every transaction) has more features to better capture the distribution of values. Our framework uses a total of 99 features for the attributes from Table \ref{tab:attributes}. Table \ref{tab:features} depicts the features used to describe the distribution of values for every attribute, where transaction is abbreviated ‘Tx’. These features (or a subset of these features) are combined into a single n-component (where n is the number of features used) Blockchain Network Structure (BNS) vector, denoted $\boldsymbol{v}$, which represents the network state over that time period. The whole process of feature processing is shown in Figure \ref{fig:aggrprocess}. Every attribute is captured by features (according to summary statistics) represented by the dots. The features are then combined into a BNS vector which represents the state of the network over that period of time. 

\begin{table}
\centering
\caption{Features}
\label{tab:features}
\arrayrulecolor{black}
\begin{tabular}{!{\color{black}\vrule}Q!{\color{black}\vrule}L!{\color{black}\vrule}W!{\color{black}\vrule}G} 
\cline{1-3}
Type                         & Attribute                               & Features                                                                                                                                                                                                                                                               & ~  \\ 
\cline{1-3}
\multirow{7}{*}{Transaction} & \multirow{2}{*}{Tx Value}               & Proportion of the values (in BTC) of transactions within the data frame in the following bins:~                                                                                                                                                                        & ~  \\ 
\cline{3-3}
                             &                                         & \multirow{9}{0.21\textwidth}{Mean, Median, Standard Deviation, Kurtosis, Skew, 10th percentile, 25th percentile, 75th percentile, 90th percentile ~ (Note that the values for Mempool Growth and Transactions per second are single point averages per minute from Blockchain.com)} &    \\ 
\cline{2-2}
                             & Tx Size                                 &                                                                                                                                                                                                                                                                        & ~  \\ 
\cline{2-2}
                             & Tx Fee Rate                             &                                                                                                                                                                                                                                                                        & ~  \\ 
\cline{2-2}
                             & Tx Fees Paid                            &                                                                                                                                                                                                                                                                        & ~  \\ 
\cline{2-2}
                             & Tx Fee Percent                          &                                                                                                                                                                                                                                                                        & ~  \\ 
\cline{2-2}
                             & Tx/Second                               &                                                                                                                                                                                                                                                                        & ~  \\ 
\cline{1-2}
\multirow{3}{*}{Mempool}     & Mempool Size                            &                                                                                                                                                                                                                                                                        & ~  \\ 
\cline{2-2}
                             & Mempool Growth                          &                                                                                                                                                                                                                                                                        & ~  \\ 
\cline{2-2}
                             & Number of Tx in Mempool                 &                                                                                                                                                                                                                                                                        & ~  \\ 
\cline{1-3}
\multirow{2}{*}{Block}       & Tx/Block                                & \multirow{2}{0.22\textwidth}{Mean, Median, Standard Deviation}                                                                                                                                                                                                                      &    \\ 
\cline{2-2}
                             & Block Size                              &                                                                                                                                                                                                                                                                        &    \\ 
\cline{1-3}
Transaction                  & Percentage of Non-Standard Transactions & \multirow{2}{0.22\textwidth}{Total value over the time period (single point)}                                                                                                                                                                                                       &    \\ 
\cline{1-2}
Address                      & Number of Unique Active Addresses       &                                                                                                                                                                                                                                                                        &    \\
\cline{1-3}
\end{tabular}
\arrayrulecolor{black}
\end{table}

In our preliminary study, we used 1,122 features which represented 27 attributes. However, we found that a high feature count per attribute (more summary statistics such as Max, Min, 5th percentile, etc.) introduced noisy variation that was highly variable and largely uncorrelated with any events. Furthermore, there was a lack of data for certain attributes at the data frequency desired. After feature selection, this study uses 99 features which represent 13 attributes. 

We found that there was too much noisy variation when using all 99 features to describe the overall Bitcoin network. Thus, we defined the Overall Bitcoin Blockchain Network Feature Set, which is a reduced set of features to describe the overall Blockchain Bitcoin Network. 

\textbf{Overall Bitcoin Blockchain Network Feature Set:} A single feature (mean or total count/value) is used per attribute. All other features are omitted except that all bins of the feature Distribution of Transaction Values are included.

The three subnetworks (activity, fee, and transaction) are captured by the feature sets shown in Table \ref{tab:subnetworks}. All features are used for each attribute. For example, for a BNS vector with $l_{df}=48$ hrs over the Transaction Network feature set, transaction-related attributes of blockchain and network data within those 48 hours would be aggregated into a single 37-element (37 features) vector that describes the Bitcoin Transaction Subnetwork for that time period.

\begin{table*}[]
\caption{Subnetworks}
\label{tab:subnetworks}
\begin{tabulary}{\linewidth}{lll}
\hline
\textbf{Subnetwork} & \textbf{Attributes} & \textbf{Features} \\ \hline
Activity Network & \begin{tabular}[c]{@{}l@{}}Block Size, Unique Addresses, Tx per second, \\ Mempool Count, Mempool Growth, Mempool Size\end{tabular} & 40 \\
Transaction Network & Distribution of Tx Values, Tx Size, Tx per second, Tx Value & 37 \\
Fee Network & Fee Rate, Total Tx Fees Paid, Tx Fee Percent & 27
\end{tabulary}
\end{table*}

We define a BNS vector and the data it captures as the following:
\begin{equation}
\begin{split}
    \boldsymbol{v}_{\boldsymbol{i}} \cong \Gamma\left[\Phi\left(\boldsymbol{v}_{\boldsymbol{i}}\right), \Omega\left(\boldsymbol{v}_{\boldsymbol{i}}\right)\right)
\end{split}
\end{equation}
where $\Phi\left(\boldsymbol{v}_{\boldsymbol{i}}\right)$ and $\Omega\left(\boldsymbol{v}_{\boldsymbol{i}}\right)$ denote the start and end time of the BNS vector respectively. $\Gamma$ denotes the blockchain/network data captured from the time interval  $\left[\Phi\left(\boldsymbol{v}_{\boldsymbol{i}}\right), \Omega\left(\boldsymbol{v}_{\boldsymbol{i}}\right)\right)$. $\Omega\left(\boldsymbol{v}_{\boldsymbol{i}}\right)=\Phi\left(\boldsymbol{v}_{\boldsymbol{i}}\right)+l_{d f}$ where length $l_{df}$ is the length of time over which the blockchain network is captured. $i$ is the order of the vector in the chronological set. In a BNS vector, each element is a feature value. 

\subsection{Impact Score}
\label{sec:iscore}
The motivation for the I-score is to provide a efficient framework to analyze the impact of various events on the Bitcoin blockchain network without computationally expensive graph matching/comparisons. The I-score provides a quantitative measure to separate day-to-day "normal" fluctuations in Bicoin blockchain network features from changes which we believe are induced by major events. Finally, the I-score provides a universal measurement of event-impact across subnetworks and configurations by standardizing changes to the baseline for that particular configuration. 

\subsubsection{I-Score Process}
\begin{figure*}
\includegraphics[width=\textwidth]{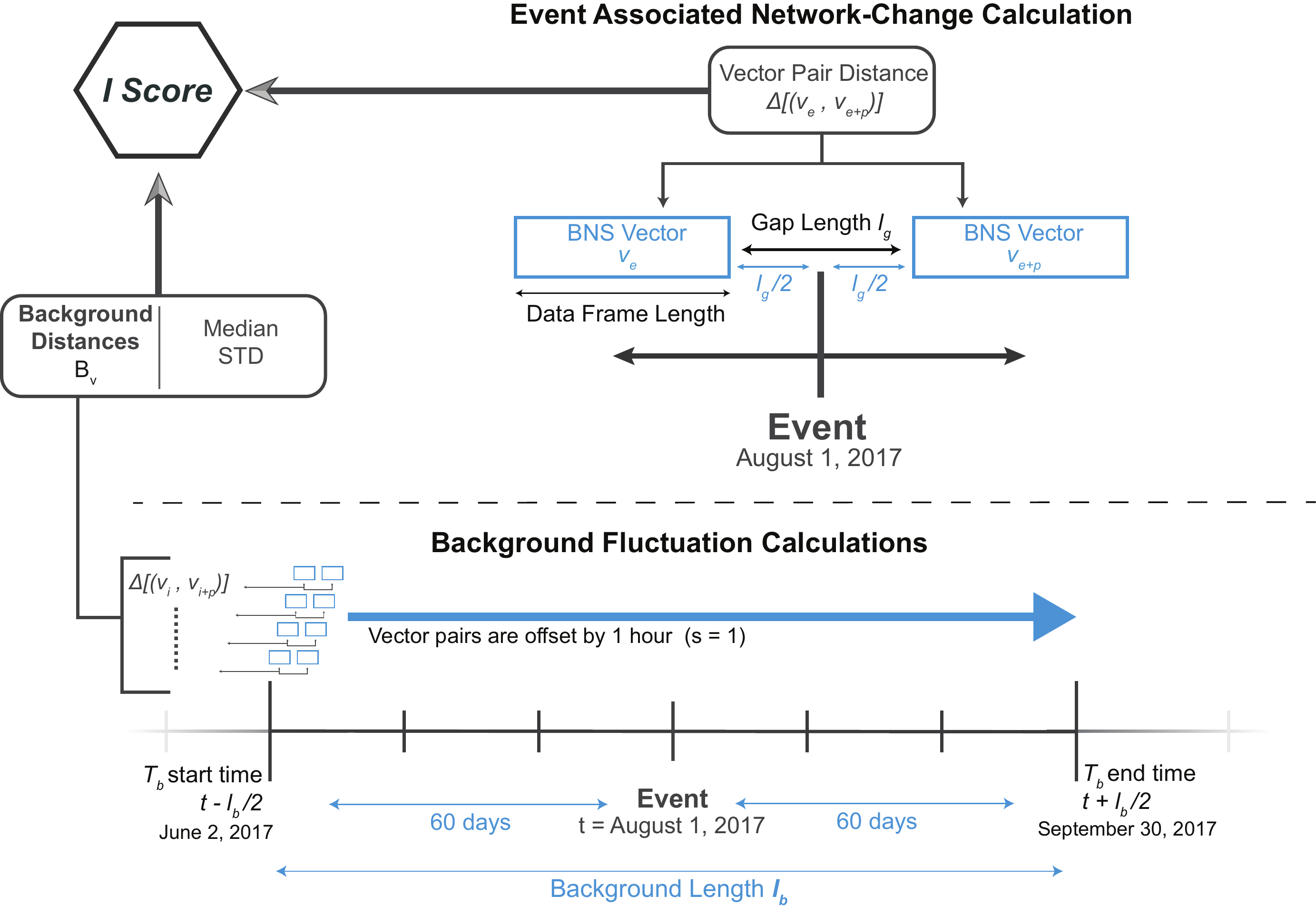}
\caption{I-Score calculation for a given event }
\label{fig:iscore}   
\end{figure*}

The blockchain network is constantly changing and has natural fluctuations, thus the blockchain network change (distance value) at a time associated with an event must be compared against background or “assumed normal” changes (distance values). We introduce the Impact Score (abbreviated I-Score) metric, which provides an indication of the significance of a network change by comparing it against day-to-day change. Figure \ref{fig:iscore} depicts the calculation of I-Score. First, we define several parameters related to the measurement of event-induced change such as event time, background length and rolling data frame overlap. Next, we measure background fluctuations by finding the average change (distance) between a large numbers of network states within the background time length. This is done by capturing a series of network states in a rolling fashion, pairing them according to the gap length specified, and calculating the change within each pair. Then, we calculate the impact score by comparing the change of the network states before/after the time of the event against the background fluctuations. 

\paragraph{Parameters}
To determine the I score of an event, the following parameters must be defined:
\begin{enumerate}
  \item Event Time = $t$: The time of the event.
  \item Data Frame Length = $l_{df} > 0$: the length in hours of the blockchain network structure to be captured into a BNS vector. For our case study, we used $l_{df}=96$ hours.
  \item Gap Length $l_{g}>0$: The length in hours between blockchain network structures. If the blockchain network change 4 hours after an event is the target of analysis, a gap length of 8 hours would be used to compare the blockchain network change from 4 hours before to 4 hours after the event.
  \item Background length = $l_{b}$: The length in days before and after the event time $t$. From this, the Background Time Period = $T={b}$, is defined over the large time interval $(t - \frac{l_{b}}{2}, t + \frac{l_{b}}{2})$. For our case study $l_{b} = 120$ days (60 days before and after each event). The exception to this is the Brexit event (see Table \ref{tab:events} in which a background length of 110 days was used due to data constraints.
  \item Rolling Data Frame Overlap = $o=(l_{df}-s)$: The overlap in hours of the network state captured of two adjacent BNS vectors in the background fluctuation calculations. In our case study, we used , $s=1$ hour, $o=95$ hours. A negative $o$ value is acceptable and indicates that the offset between two adjacent vectors is greater than their data frame length. 
\end{enumerate}
For proper and easy pairing of BNS vectors in the I-Score calculation, we recommend Rolling Data Frame Overlap = $l_{df}-1$, $s=1$ hour.

\paragraph{Background Fluctuation Calculations}
We define $E_{v}$ as a series of network states (BNS vectors) in chronological order within $T_{b}$. 
\begin{equation}
\begin{split}
    E_{v}=\left\langle v_{0}, \dots, v_{i} \ldots, v_{n}\right\rangle
\end{split}
\end{equation}

Where each $v_{i}$ is defined as follows:

\begin{equation}
\begin{split}
    v_{0} \cong \Gamma\left[\Phi\left(v_{0}\right), \Omega\left(v_{0}\right)\right), \\ \Phi\left(v_{0}\right)=t-\frac{l_{b}}{2} \text { and } \Omega\left(v_{0}\right)=t-\frac{l_{b}}{2}+l_{d f}
\end{split}
\end{equation}

\begin{equation}
\begin{split}
    \boldsymbol{v}_{\boldsymbol{i}} \cong \Gamma\left[\Phi\left(\boldsymbol{v}_{\boldsymbol{i}-\mathbf{1}}\right)+s, \Omega\left(\boldsymbol{v}_{\boldsymbol{i}-\mathbf{1}}\right)+s\right) \\ \text { for } 0<i<n
\end{split}
\end{equation}

\begin{equation}
\begin{split}
    v_{n} \cong \Gamma\left[\Phi\left(v_{n}\right), \Omega\left(v_{n}\right)\right), \\ \Phi\left(v_{n}\right)=t+\frac{l_{b}}{2}-l_{d f} \text { and } \Omega\left(v_{n}\right)=t+\frac{l_{b}}{2}
\end{split}
\end{equation}

BNS vectors from $E_{v}$ are first paired and added to the set of all network changes. A BNS vector pair is defined as the pair of two BNS vectors $\left(\boldsymbol{v}_{i}, \boldsymbol{v}_{\boldsymbol{i}+\boldsymbol{p}}\right)$ where $\Phi\left(\mathbf{v}_{i+p}\right)- \Omega\left(\mathbf{v}_{i}\right)=l_{g}$. BNS vector pairs are then added to the set of all changes which is defined as follows:

\begin{equation}
C_{v}=\left\{\left(\boldsymbol{v}_{0}, \boldsymbol{v}_{\mathbf{0}+\boldsymbol{p}}\right), \ldots,\left(\boldsymbol{v}_{\boldsymbol{n}-\boldsymbol{p}} \boldsymbol{v}_{\boldsymbol{n}}\right)\right\}
\end{equation}

 The distance value of each pair in $C_{v}$ is calculated using the distance measure specified (represented by $\Delta\left[\left(\boldsymbol{v}_{i}, \boldsymbol{v}_{\boldsymbol{i}+\boldsymbol{p}}\right)\right]$) and added to the set of all background distance values. 
 \begin{equation}
     B_{d}=\left\{\Delta\left[\left(\boldsymbol{v}_{i}, \boldsymbol{v}_{i+\boldsymbol{p}}\right)\right] |\left(\boldsymbol{v}_{i}, \boldsymbol{v}_{\boldsymbol{i}+\boldsymbol{p}}\right) \in C_{v}\right\}
 \end{equation}

In our case study, we used Squared Euclidean Distance to measure the change between the BNS vectors. BNS vectors were standardized using the MinMaxScaler from Scikit-learn \cite{pedregosa2011scikit} to $(0,1)$. Squared Euclidean Distance is defined as follows.
\begin{multline}
    d(p, q)=\left(p_{1}-q_{1}\right)^{2}+\cdots+\left(p_{i}-q_{i}\right)^{2}+\cdots+\left(p_{n}-q_{n}\right)^{2}
\end{multline}
where $p$ and $q$ are BNS vectors.

A larger Squared Euclidean Distance value between two BNS vectors indicates a greater change between the state of the Bitcoin blockchain network/subnetwork captured by those BNS vectors. Five distance measures were initially used to compare vectors: Euclidean Distance, Squared Euclidean Distance, Cosine Distance, Canberra Distance and Chebyshev Distance. After preliminary testing, we determined that Squared Euclidean Distance was the most suited due to the squaring of each dimension which allows for easier separation of event change and background fluctuations

\paragraph{Event Associated Network Change}
The I-Score is defined as follows:
\begin{equation}
    \mathrm{I}=\frac{\Delta\left[\left(\boldsymbol{v}_{e}, \boldsymbol{v}_{e+\boldsymbol{p}}\right)\right]-M D\left(B_{d}\right)}{\operatorname{SD}\left(B_{d}\right)}
\end{equation}
Where $MD(B_{d})$ and $SD(B_{d})$ denote the median and standard deviation of the set $B_{d}$ respectively. \(\boldsymbol{v}_{\boldsymbol{e}}\) and \(\boldsymbol{v}_{\boldsymbol{e}+\boldsymbol{p}}\) denote the network states before and after the event respectively. 

\begin{equation}
    \begin{split}
        \boldsymbol{v}_{e} \cong \Gamma\left[\Phi\left(\boldsymbol{v}_{e}\right), \Omega\left(\boldsymbol{v}_{e}\right)\right), \\ \Phi\left(\boldsymbol{v}_{e}\right)=t-\frac{l_{g}}{2}-l_{d f} 
        \text { and } \Omega\left(\boldsymbol{v}_{e}\right)=t-\frac{l_{g}}{2}
    \end{split}
\end{equation}
\begin{equation}
    \begin{split}
        \boldsymbol{v}_{\boldsymbol{e}+\boldsymbol{p}} \cong \Gamma\left[\Phi\left(\boldsymbol{v}_{\boldsymbol{e}+\boldsymbol{p}}\right), \Omega\left(\boldsymbol{v}_{\boldsymbol{e}+\boldsymbol{p}}\right)\right), \\ \Phi\left(\boldsymbol{v}_{\boldsymbol{e}+\boldsymbol{p}}\right)=t+\frac{l_{g}}{2}  \text { and } \Omega\left(\boldsymbol{v}_{\boldsymbol{e}+\boldsymbol{p}}\right)=t+\frac{l_{g}}{2}+l_{d f}
    \end{split}
\end{equation}

The numerical result (introduced in Section \ref{sec:iscore}) of the I-score process is derived from the Z-score and the modified Z-score. The Z score is a measure of standard deviations above/below a population mean that used in statistics. The modified Z score uses the Median Absolute Deviation (MAD) and the median in place of standard deviation and mean. In the I score, we use a hybrid of the Z-score and the modified Z-score by using the median with standard deviation. 

\subsubsection{Time Complexity}
The time complexity of our Impact Score algorithm is \(O\left(m * n^{2}\right)\) where $n$ is the number of BNS vectors in $T_b$ and $m$ is the number of features used in each BNS vector. If overlap $o = l_{df} -1$, the time complexity of the algorithm reduces to \(O\left(m * n\right)\) as the second vector \(\boldsymbol{v}_{\boldsymbol{i} + \boldsymbol{p}}\) in each vector pair \(\left(\boldsymbol{v}_{i}, \boldsymbol{v}_{\boldsymbol{i}+\boldsymbol{p}}\right)\) can be found in constant time complexity using \(p=\frac{l_{d f}+l_{g}}{s}-1\). On a 12-core 2.0 Ghz, 128 GB RAM Windows VM, the average time performance (tested over 50 trials) of our impact score algorithm for a single event with parameters [$T_{b}=120$ days, $l_{df}=96$ and $o=95$] is 7.128 seconds.

\subsubsection{Advantages of the I-Score}
Our I-Score metric offers two distinct advantages over distance values alone for measuring event-impact.

First, the I-Score metric is a universal measurement of event-impact across all configurations of blockchain networks, subnetworks, gap lengths, and data frame lengths within a cryptocurrency. It allows us to directly compare all event-induced changes – such as a gap length of 0 hours vs 200 hours, overall network vs fee subnetwork – as the metric standardizes event-induced changes to the natural fluctuations for that given configuration. Such a comparison cannot be made using the distance value directly as distance between network structure vectors increases proportional to the gap length (greater time between network states = more change), and subnetworks and data frame lengths have differing distance scales, natural fluctuations etc. Thus, by standardizing event distance to background fluctuation distances for the given configuration, the I-Score metric allows us to make direct comparisons across all configurations. 

Second, The I-Score metric provides an indication of event impact significance by separating event induced network changes from natural fluctuations. This metric is robust across all configurations as the event change is compared to the natural fluctuations of the given configuration. 

\subsubsection{Limitations of the I-Score}
A limitation in our I-Score metric is its use of a background time period associated with the time of the event. As such, the I-Score metric is the best indicator of event significance when there are relatively few events in the background time period. During extremely eventful periods of time, such as late-2017 to early-2018 for our case study on Bitcoin, the I-Score for a given event underestimates the significance of any event-induced change. A further discussion of the I-score is presented in Section \ref{sec:IScoreDiscussion}.

\section{Case Study}
\label{sec:casestudy}
Our case study aims to answer the following research questions:
\begin{enumerate}
  \item Do external and blockchain-intrinisic events cause an observable change on the Bitcoin blockchain network or its subnetworks?
  \item Is there a link between event type and impact: temporal, subnetwork, or network change? Are event-associated temporal impacts discernable by type?
  \item Are the event-induced changes observed consistent with expectations? Are there unexpected similarities between events and their impact?
\end{enumerate}

\subsection{Setup}
\label{sec:setup}
To evaluate our framework, we have selected 16 events occurring between 2016-2018, as listed in Table \ref{tab:events}. We believed that they have induced varying levels and forms of change on the Bitcoin blockchain network structure. We have studied each event with the overall feature set and the reduced feature sets at gap lengths between 0-480 hours to analyze impact on the overall Bitcoin blockchain network and its subnetworks. The events fall into three groups: 

\begin{enumerate}
    \item \textbf{Global} (denoted G) – Security, political, or governmental events that impact the Bitcoin ecosystem either directly or indirectly.
    \item \textbf{Financial} (denoted F) – Events that impact or are the result of a change in the state or value of Bitcoin as a cryptocurrency.
    \item \textbf{Regulatory} (denoted R) – Events that are the result of either positive or negative regulation directly or indirectly affecting the Bitcoin ecosystem. 
\end{enumerate}

Our process is not able to associate an event with an observed impact with 100\% certainty, as our process measures the change relative to fluctuations at the time of an event. Thus, we were careful to select 16 events with few other major events around the same event time. Despite the presence of other events that have also affected the network, we believe the select event was the most significant at the time. 

Due to the rather eventful nature of Bitcoin in our time period of study, we considered an event to have a substantial impact if its I-Score was greater than $> 1.9$. The I-Score metric generally underestimates the “true” impact of an event due to the presence of other events during the background time period. Initially, the impact threshold was set at I-Score $> 2.0$, however after testing we found that the I-Score for a given event fluctuated approximately $\pm 0.1$ if the event time was adjusted between $\pm 3$ hours. Since the event time of certain events were approximate (i.e. there is no exact time when the Brexit referendum was announced due to various networks forecasting at different rates), we set the substantial impact threshold at 1.9. Furthermore, we set a significant impact threshold at I-Score $> 2.9$.

\newcolumntype{Z}{>{\arraybackslash}m{10cm}}
\begin{table*}
\centering
\caption{Events in Chronological Order}
\label{tab:events}
\arrayrulecolor{black}
\begin{tabular}{|M!{\color{black}\vrule}W|Z|} 
\arrayrulecolor{black}\hline
\textbf{ N \& Type } & \textbf{ Event Name }                & \textbf{ Details  }                                                                                                                                         \\ 
\hline
1G                   & Brexit (June 2016)                   & UK votes to leave the European Union \cite{bovaird_bovaird_2016_brexitbitfinex}.                                                                                                                       \\ 
\arrayrulecolor{black}\hline
2G                   & Bitfinex Hacked (Aug 2016)           & Bitfinex (the largest Bitcoin exchange at that time) halts all trading. They later announce that 119,756 bitcoins have been stolen in a security incident \cite{bovaird_bovaird_2016_brexitbitfinex, bitfinex}.  \\ 
\hline
3G                   & Trump Elected (Nov 2016)             & Donald Trump is elected president of the United States \cite{fortune_trumpvictory}.                                                                                                     \\ 
\hline
4F                   & 1 BTC=\$1000 USD (Jan 2017)          & The exchange rate of Bitcoin and USD to hits 1 BTC = \$1000 for the second time in Bitcoin history since 2014.                                              \\ 
\hline
5R                   & SEC denies ETF (March 2017)          & The SEC denies the Winklevoss twins’ application to operate a Bitcoin ETF \cite{michaels_2018}. This is the first Bitcoin ETF application.                                       \\ 
\hline
6R                   & Japan legalizes BTC (April 2017)     & Bitcoin is officially accepted as a legal payment method in Japan starting April 1, 2017 \cite{garber_2017}.                                                                   \\ 
\hline
7G                   & BTC-BCH Hard Fork (Aug 2017)         & Bitcoin is hard forked to create Bitcoin Cash (BCH), a new cryptocurrency with its own blockchain in addition to Bitcoin \cite{lukewgraham_2017}.                                   \\ 
\hline
8F                   & CME Announces BTC Futures (Oct 2017) & CME Group announces that it will launch Bitcoin futures in Q4 of 2017 \cite{futures}                                                                                       \\ 
\hline
9F                   & 1 BTC = \$10,000 USD (Nov 2017)      & The exchange rate of Bitcoin and USD hits 1 BTC = \$10,000 for the first time in Bitcoin history.                                                           \\ 
\hline
10F                  & CBOE BTC Futures Launch (Dec 2017)   & The CBOE futures exchange launches the first ever Bitcoin futures contract \cite{products}.                                                                                 \\ 
\hline
11F                  & BTC Price Peak (Dec 2017)            & The exchange rate of Bitcoin and USD reaches a record high of 1 BTC = \$19,783.06.                                                                          \\ 
\hline
12R                  & South Korea BTC Ban (Jan 2018)       & The South Korean government announces a possible ban on Bitcoin trading \cite{zhong_2017}.                                                                                    \\ 
\hline
13G                  & Coincheck NEM Hack (Jan 2018)        & The Coincheck cryptocurrency exchange is hacked. Hackers steal over 500 million dollars in NEM coins (a different cryptocurrency) \cite{bloomberg_2018}.                          \\ 
\hline
14R                  & Facebook bans Crypto Ads (Jan 2018)  & Facebook announces a ban on cryptocurrency ads on its platform \cite{facebook_business}.                                                                                             \\ 
\hline
15R                  & SEC Exchange Register (March 2018)   & The SEC announces that all cryptocurrency exchanges must register with the SEC \cite{sec}.                                                                             \\ 
\hline
16R                  & Google bans Crypto Ads (March 2018)  & Google announces a ban on all cryptocurrency-related ads \cite{spencer_2018}.                                                                                                   \\
\arrayrulecolor{black}\hline
\end{tabular}
\end{table*}

For Bitcoin blockchain data, we obtained an API key from Blockchain.com \cite{blockchain} (formerly Blockchain.info) which was used to download Blocks 0-534,400 of the Bitcoin blockchain into a MongoDB database for analysis. Mempool Size, Mempool Count, Mempool Growth, and Transaction Rate data \textemdash which are not recorded on the Bitcoin blockchain \textemdash was downloaded from Blockchain.com Charts \cite{blockchain}. Due to the absence of Mempool and Transaction rate data before April 24, 2016, our case study analyzed the impact of major events that occurred after April 24, 2016. We created a custom Java program to create BNS vectors from Bitcoin blockchain and network data and used Scikit-learn \cite{pedregosa2011scikit} in Python for our Impact-Score analysis process. Matplotlib was used to create figures \cite{Hunter:2007}. 

\subsection{Results}
\label{sec:results}
Of the 16 events, 15 were associated with a substantial (I-Score $>$ 1.9) change in the overall Bitcoin blockchain network or one or more of its subnetworks. The one event that was not associated with any substantial change in was the legalization of BTC by Japan on April 1, 2017 (Event 6). Among all 16 events, the election of Trump (Event 3) resulted in the greatest network change with an I-Score of 5.459 at 60 hours after the event.  

\subsection{Event-Induced Changes in the Overall Bitcoin Blockchain Network}
\label{sec:OverallBBN}
9 of the 16 events were associated with a substantial (I-Score $>$ 1.9) change in the overall Bitcoin blockchain network (defined over our \textit{Overall Bitcoin Blockchain Network Feature Set}). Furthermore 14 of the 16 events were associated with a discernable (I-Score $>$ 1.0) change in the overall Bitcoin blockchain network. 
Note that the I-Scores of the South Korea BTC Ban, Coincheck NEM Hack, CBOE BTC Futures Launch, BTC Price Peak, Facebook bans Crypto Ads and 1 BTC =\$10k events (highlighted in blue) significantly underestimates their true impact due to their extremely eventful background time periods. From October 2017 - February 2018 there was rampant speculation, “Bitcoin hype”, increased regulatory attention from countries all around the world, dozens of high-profile Bitcoin events from the launch of Bitcoin futures to the debut of Bitcoin Gold, and widespread media coverage during this time. As a testament to public frenzy, in November 2017, the Bitcoin exchange Coinbase added over 300,000 new users in a single week and doubled their user base from 2016 \cite{chengevelyn_2017_coinbaseusers}.

\begin{table}
\caption{Event Impact on the Overall Bitcoin Blockchain}
\centering
\arrayrulecolor{black}
\begin{tabular}{!{\color{black}\vrule}S!{\color{black}\vrule}W!{\color{black}\vrule}M!{\color{black}\vrule}M!{\color{black}\vrule}} 
\hline
N \& Type & Event                     & Max I-Score                          & Temporal Delay  \\ 
\hline
3G        & Trump Elected             & 5.46                                 & 60              \\ 
\hline
16R       & Google bans Crypto Ads    & 3.93                                 & 5               \\ 
\hline
7G        & BTC-BCH Hard Fork         & 3.58                                 & 0               \\ 
\hline
1G        & Brexit                    & 3.39                                 & 168             \\ 
\hline
11F       & BTC Price Peak            & {\cellcolor[rgb]{0.624,0.875,1}}3.25 & 84              \\ 
\hline
5R        & SEC denies ETF            & 2.32                                 & 192             \\ 
\hline
10F       & CBOE BTC Futures Launch   & {\cellcolor[rgb]{0.624,0.875,1}}2.29 & 192             \\ 
\hline
8F        & CME Announces BTC Futures & 2.07                                 & 240             \\ 
\hline
14R       & Facebook bans Crypto Ads  & {\cellcolor[rgb]{0.624,0.875,1}}1.97 & 240             \\ 
\hline
13G       & Coincheck NEM Hack        & {\cellcolor[rgb]{0.624,0.875,1}}1.33 & 20              \\ 
\hline
4F        & 1 BTC=\$1000 USD          & 1.22                                 & 13              \\ 
\hline
9F        & 1 BTC = \$10k USD         & {\cellcolor[rgb]{0.624,0.875,1}}1.2  & 15              \\ 
\hline
15R       & SEC Exchange Register     & 1.04                                 & 144             \\ 
\hline
6R        & Japan legalizes BTC       & 1                                    & 108             \\ 
\hline
12R       & South Korea BTC Ban       & {\cellcolor[rgb]{0.624,0.875,1}}0.71 & 10              \\ 
\hline
2G        & Bitfinex Hacked           & 0.02                                 & 60              \\
\hline
\end{tabular}
\arrayrulecolor{black}
\end{table}
The temporal delay indicates the time (in hours after the event) at which the maximum I-Score occurred. 

\begin{figure*}
\includegraphics[width = 0.95\textwidth]{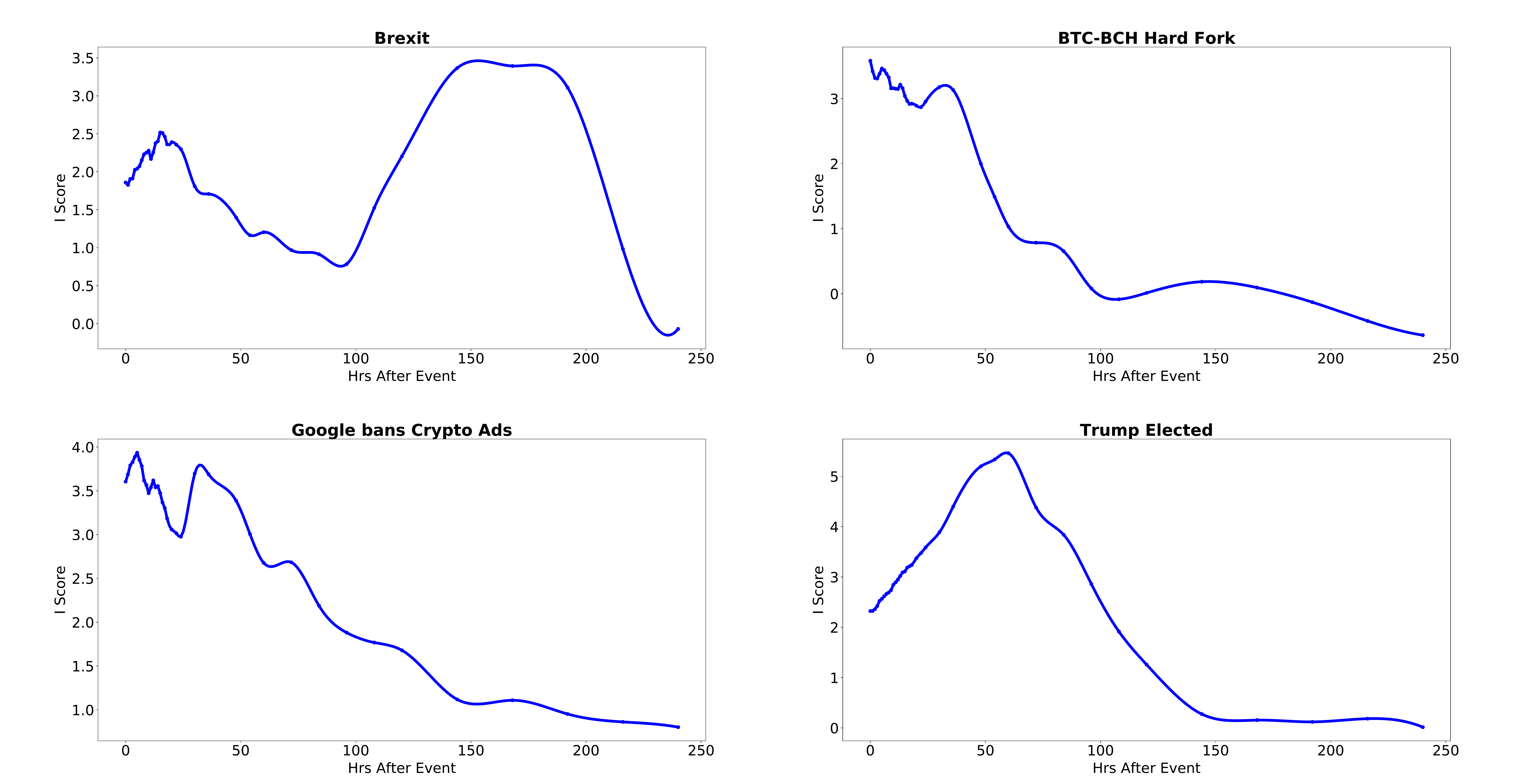}
\caption{The top four events with greatest I-Score on the Overall Bitcoin blockchain network}
\label{fig:top4}   
\end{figure*}

Looking at the five events with an I-Score greater than 3, the top four events (by I-Score) as shown in Fig. \ref{fig:top4} are associated with an immediate substantial change (starting 0 hours after the event) in the overall Bitcoin Blockchain network. The exceptional event is the BTC Price Peak event (Event 11). We believe the lack of an immediate substantial change in the overall Bitcoin blockchain network associated with the peak record price of Bitcoin is likely due to the hindsight nature of this event – that is this event was only viewed as such due to the record price looking back retrospectively and during the time of the event, there was no indication that this was the record price. 

\begin{figure}
\includegraphics[width = 0.5\textwidth]{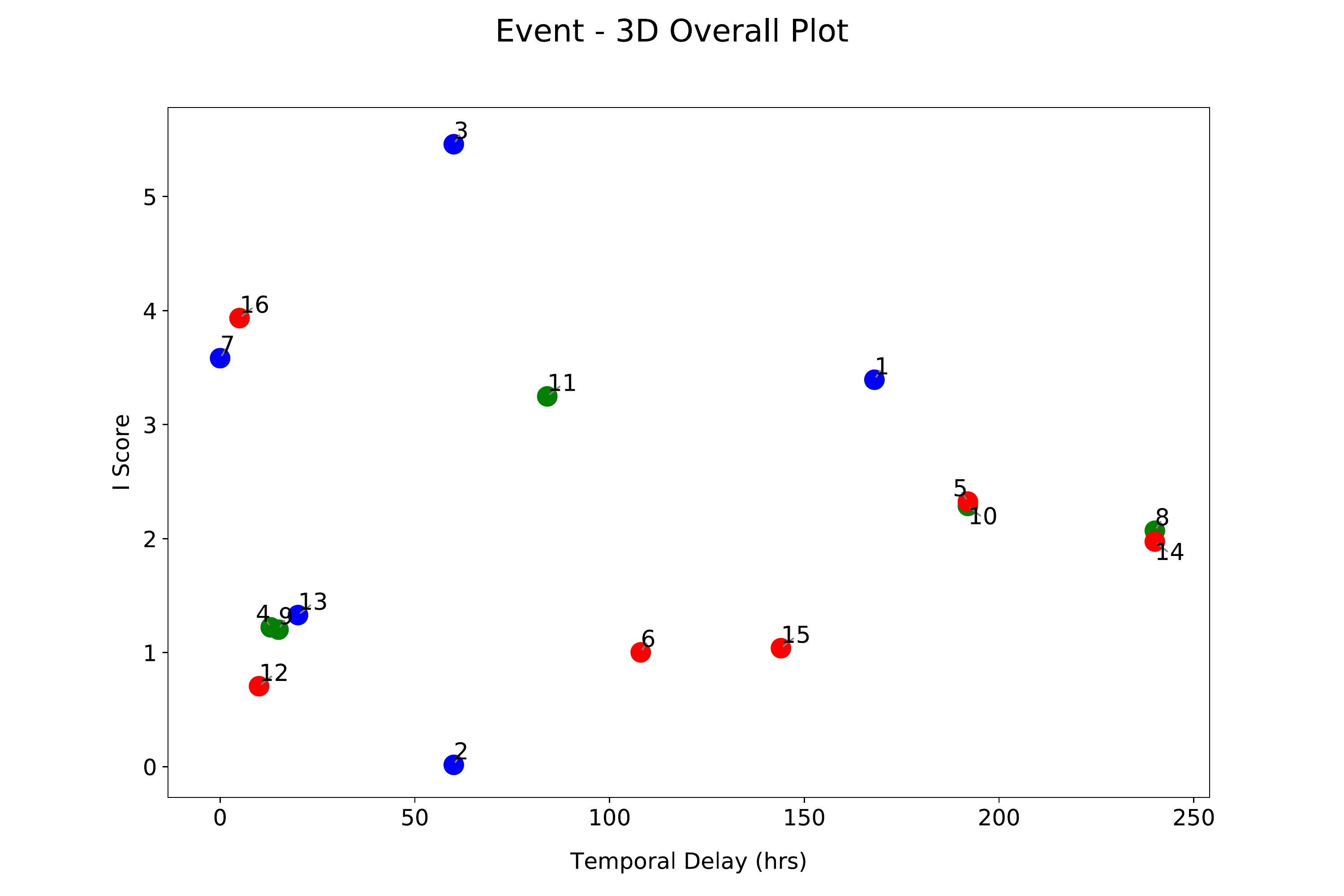}
\caption{Event Impact on the Overall Bitcoin blockchain network. Event types: Global = Blue, Financial = Green, Regulatory = Red}
\label{fig:typeandoverall}   
\end{figure}

As shown in Fig. \ref{fig:typeandoverall}, there seems to be no clear correlation between the event type and its impact, both in the temporal sense and the impact as measured by the I-Score, on the overall Bitcoin network. However, Fig. 5 also depicts similarities between individual events. One of these similarities is between the 1 BTC = \$1000 USD and 1 BTC = \$10k USD (Events 4 and 9). As shown in Fig. \ref{fig:typeandoverall} and Fig. 6, the temporal effect and overall impact of these two events show remarkable similarity. Both events represent the crossing of a major milestone in Bitcoin exchange price and it seems that the overall Bitcoin blockchain network reacts in a similar manner to both value milestones. Note that in both events, there is an initial change in the blockchain network and with a maximum change at a temporal delay of 12-16 followed by a decline in network change and a second delayed “spike” in network change at a temporal delay of 130-160 hours. We postulate that the relatively low I-Score associated with these two events is likely the result of general momentum in the weeks or days before as price rallies – thus resulting in a less drastic change when the milestone is crossed.

\begin{figure}
\includegraphics[width = 0.5\textwidth]{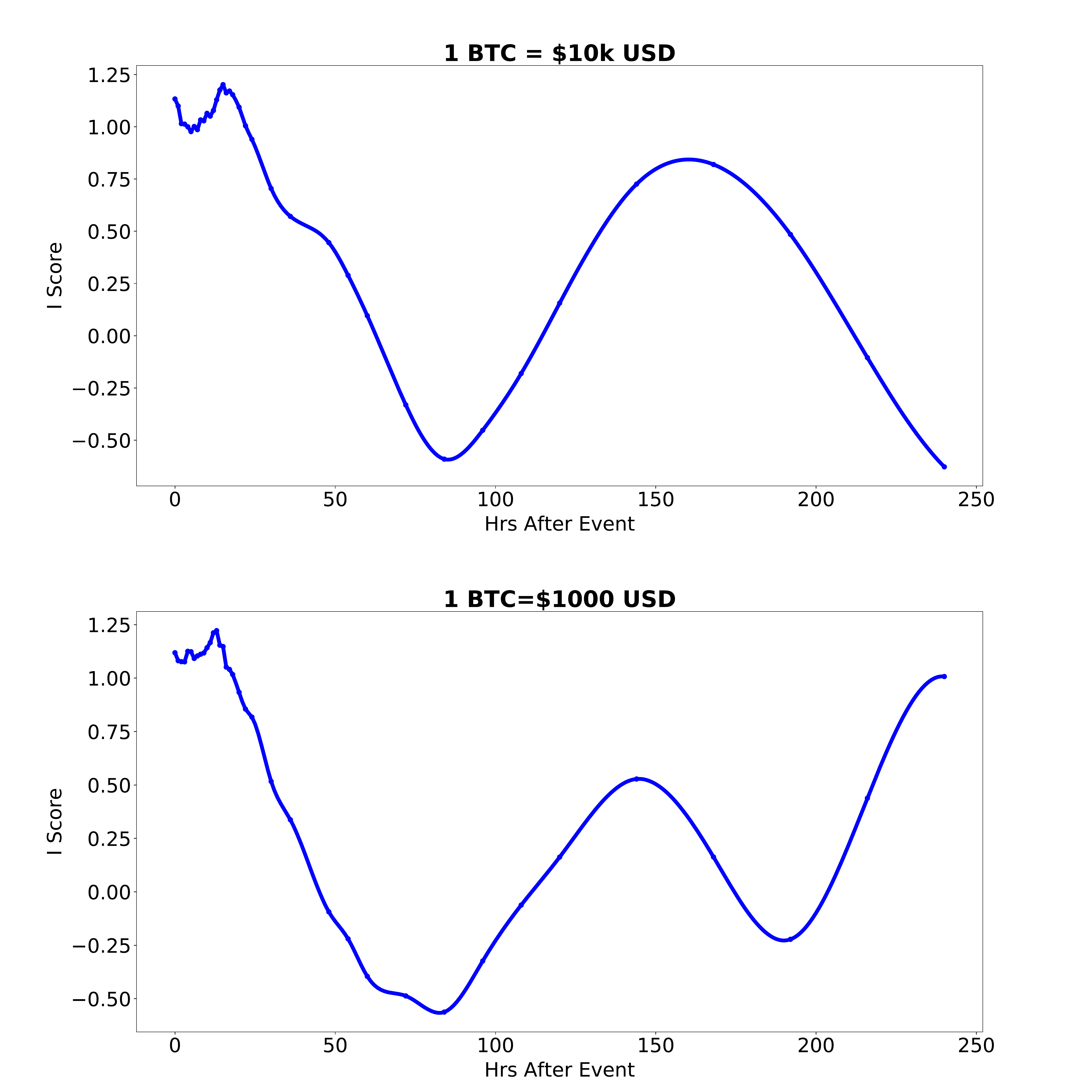}
\caption{Two exchange price ‘milestone’ events with similar temporal impacts}
\label{fig:btcmilestone}   
\end{figure}

\subsection{Event Type and Impact}
\label{sec:typeimpact}
Tables \ref{tab:global}, \ref{tab:regulatory}, and \ref{tab:financial} show the impact measured in I-Score of each type of event on the three Bitcoin subnetworks without the temporal impact component. I-Scores greater than 2.9 are shaded in red and I-Scores greater than 1.9 (defined as the threshold for substantial impact) are shaded in orange.  Transaction Network is abbreviated TxN, Activity Network is abbreviated AN, and Fee Network is abbreviated FN.

\begin{table}
\centering
\arrayrulecolor{black}
\caption{I-Scores of Global Events}
\label{tab:global}
\begin{tabular}{|l!{\color{black}\vrule}l|S|S|S|} 
\hline\hline
N  & Event                                           & Max TxN                                  & Max AN                                   & Max FN                                    \\ 
\hline
3  & Trump Elected                                   & {\cellcolor[rgb]{0.996,0.706,0.749}}3.11 & 0.24                                     & 1.41                                      \\ 
\hline
7  & BTC-BCH Hard Fork                               & {\cellcolor[rgb]{0.996,0.706,0.749}}3.93 & 0.48                                     & 0.92                                      \\ 
\hline
1  & Brexit                                          & {\cellcolor[rgb]{1,0.851,0.702}}2.90     & {\cellcolor[rgb]{0.996,0.706,0.749}}4.45 & {\cellcolor[rgb]{0.996,0.706,0.749}}4.06  \\ 
\hline
13 & Coincheck NEM Hack                              & 1.36                                     & {\cellcolor[rgb]{1,0.851,0.702}}2.62     & 0.71                                      \\ 
\hline
2  & Bitfinex Hacked                                 & 1.02                                     & 0.40                                     & {\cellcolor[rgb]{1,0.851,0.702}}2.72      \\ 
\hline
\multicolumn{2}{|l!{\color{black}\vrule}}{Average}   & {\cellcolor[rgb]{0.851,0.886,0.953}}2.46 & {\cellcolor[rgb]{0.851,0.886,0.953}}1.64 & {\cellcolor[rgb]{0.851,0.886,0.953}}1.96  \\
\hline\hline
\end{tabular}
\arrayrulecolor{black}
\end{table}

\begin{table}
\centering
\caption{I-Scores of Regulatory Events}
\label{tab:regulatory}
\arrayrulecolor{black}
\begin{tabular}{|l!{\color{black}\vrule}l|S|S|S|} 
\hline\hline
N  & Event                                                        & Max TxN                                  & Max AN                                   & Max FN                                    \\ 
\hline\hline
16 & Google bans Crypto Ads                                       & {\cellcolor[rgb]{0.996,0.706,0.749}}3.00 & {\cellcolor[rgb]{1,0.851,0.702}}2.31     & {\cellcolor[rgb]{0.996,0.706,0.749}}3.26  \\ 
\hline
5  & SEC denies ETF                                               & 1.72                                     & 1.12                                     & {\cellcolor[rgb]{1,0.851,0.702}}2.81      \\ 
\hline
14 & Facebook bans Crypto Ads                                     & 1.13                                     & {\cellcolor[rgb]{1,0.851,0.702}}2.87     & 1.88                                      \\ 
\hline
15 & SEC Exchange Register                                        & {\cellcolor[rgb]{0.996,0.706,0.749}}3.00 & {\cellcolor[rgb]{1,0.851,0.702}}2.03     & {\cellcolor[rgb]{0.996,0.706,0.749}}4.22  \\ 
\hline
12 & South Korea BTC Ban                                          & {\cellcolor[rgb]{0.996,0.706,0.749}}3.04 & {\cellcolor[rgb]{1,0.851,0.702}}2.94     & 0.08                                      \\ 
\hline
\multicolumn{2}{|l!{\color{black}\vrule}}{Average (No Event 6)}   & {\cellcolor[rgb]{0.851,0.886,0.953}}2.38 & {\cellcolor[rgb]{0.851,0.886,0.953}}2.25 & {\cellcolor[rgb]{0.851,0.886,0.953}}2.45  \\ 
\hline
6  & Japan legalizes BTC                                          & 0.51                                     & 0.03                                     & 0.00                                      \\
\hline\hline
\end{tabular}
\arrayrulecolor{black}
\end{table}

\begin{table}
\centering
\caption{I-Scores of Financial Events}
\label{tab:financial}
\arrayrulecolor{black}
\begin{tabular}{|l!{\color{black}\vrule}l|S|S|S|} 
\hline\hline
N  & Event                                           & Max TxN                                  & Max AN                                   & Max FN                                        \\ 
\hline
11 & BTC Price Peak                                  & {\cellcolor[rgb]{1,0.851,0.702}}1.93     & 0.58                                     & {\cellcolor[rgb]{0.996,0.706,0.749}}4.36      \\ 
\hline
10 & CBOE BTC Futures Launch                         & 1.77                                     & 1.07                                     & {\cellcolor[rgb]{0.749,0.749,0.749}}4.35 (0)  \\ 
\hline
8  & CME Announces BTC Futures                       & 1.12                                     & {\cellcolor[rgb]{0.996,0.706,0.749}}3.19 & 1.93                                          \\ 
\hline
4  & 1 BTC=\$1000 USD                                & {\cellcolor[rgb]{1,0.851,0.702}}2.18     & {\cellcolor[rgb]{1,0.851,0.702}}2.83     & {\cellcolor[rgb]{0.996,0.706,0.749}}3.64      \\ 
\hline
9  & 1 BTC = \$10k USD                               & 1.85                                     & {\cellcolor[rgb]{1,0.851,0.702}}1.96     & 0.07                                          \\ 
\hline
\multicolumn{2}{|l!{\color{black}\vrule}}{Average}   & {\cellcolor[rgb]{0.851,0.886,0.953}}1.77 & {\cellcolor[rgb]{0.851,0.886,0.953}}1.93 & {\cellcolor[rgb]{0.851,0.886,0.953}}2.00      \\
\hline\hline
\end{tabular}
\arrayrulecolor{black}
\end{table}

As shown in Table \ref{tab:global}, our data suggests that global events generally impact the transaction subnetwork to a moderate degree (note the wide range in I-Score) and impact the activity and fee subnetworks to a varying extent. By splitting the global events into subgroups, we observe that the two political events not intrinsically tied to Bitcoin, Brexit (Event 1) and Trump Elected (Event 3), had a significant and substantial impact on the Bitcoin transaction subnetwork. We postulate that major global events that induce FUD (fear, uncertainty, and doubt) in traditional government/economic systems substantially boost the attractiveness of Bitcoin (given it’s decentralized “safe-haven” status), thus impacting the Bitcoin transaction subnetwork. Furthermore, we observe that the two Bitcoin security-type events, Coincheck NEM Hack (Event 13) and Bitfinex Hack (Event 2), did not substantially impact the Bitcoin transaction network. We postulate that Bitcoin or cryptocurrency security-related incidents do not substantially alter the perspective and outlook of Bitcoin users.

As shown in Table \ref{tab:regulatory}, our data suggests that regulatory events (ignoring the outlier of Event 6) impact all three subnetworks to a moderate degree. We observe a generally consistent substantial impact on the activity subnetwork and a highly variable impact on the fee subnetwork. We believe that this variable impact is due to the wide range of regulation events (some are directly Bitcoin related (SK BTC Ban) while others are indirectly related (Google bans Crypto Ads)).  

As shown in Table \ref{tab:financial}, we observe no strongly generalizable impact on the Bitcoin subnetworks as a result of financial type events. However, our data suggests that all financial events impact the Bitcoin transaction network to a moderate extent which is consistent with expected behavior. Note that the Fee I-Score of Event 10 is grayed out. This is because the I-Score CBOE BTC Futures Launch on the fee subnetwork is inaccurate. 
\begin{figure}
\includegraphics[width = 0.5\textwidth]{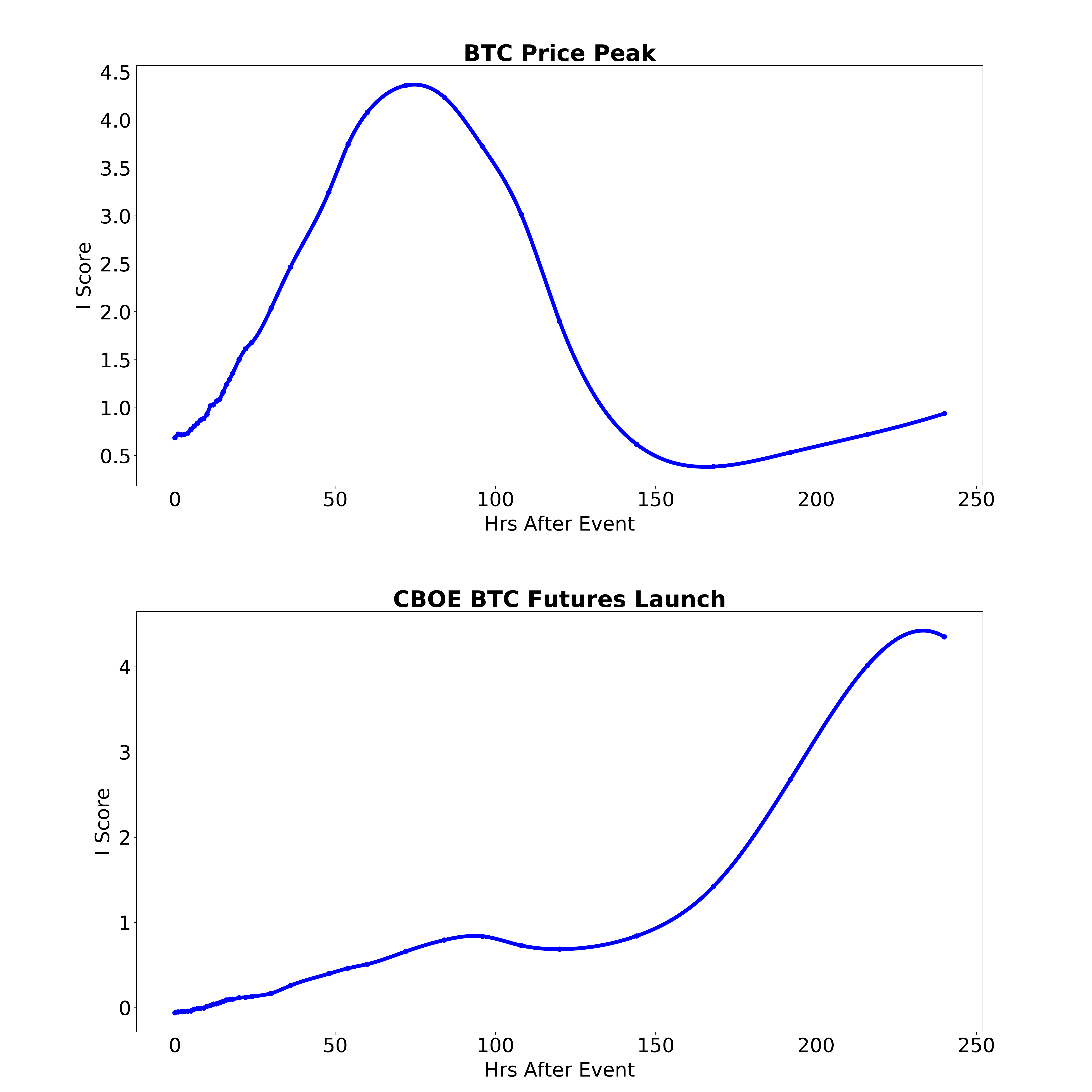}
\caption{CBOE BTC Futures Launch and BTC Price Peak overlap}
\label{fig:cboepeak}   
\end{figure}
Due to the 7-day separation between the CBOE BTC Futures Launch (December 10) and the BTC Price Peak (December 17), the analysis periods overlap. Thus, the fee subnetwork change associated with the CBOE BTC Futures Launch 240 hours after this event (note the extremely similar I-Scores between the two events) is a false association – this fee subnetwork change is the fee subnetwork change 72 hours after the BTC Price Peak. As shown in the temporal fee network change graphs in Fig. \ref{fig:cboepeak}, the same peak can be seen on the temporal graphs of both events and are separated by 240-72 = 168 hours or 7 days – exactly the time between the two events. From this, we conclude that there is not a discernible change in the fee subnetwork associated with the CBOE BTC Futures Launch. The similarity in peak shapes from these two overlaps supports our usage of the cubic spline interpolation between data points in the construction of our temporal I-Score graphs.

In the following, we present in-depth analysis of three major events. 
\begin{enumerate}
  \item 2016 Election of Trump (Global Event)
  \item Bitcoin – Bitcoin Cash Hard Fork (Global Event)
  \item Proposed ban on Bitcoin announced by South Korea (Regulatory Event)
\end{enumerate}

\subsection{2016 US Election of Donald Trump}
\label{us2016}
The 2016 US Election of Trump is a global political event that is associated with the largest I-Score. Due to the decentralized, government-less nature of Bitcoin, it is viewed by some as a prime asset in times of economic instability or political uncertainty. For example, the price of Bitcoin and overall Bitcoin buzz increased during events such as the Cyprus Crisis or Brexit. In the direct aftermath of the Election of Trump, Fortune reported that futures markets dropped while gold and Bitcoin increased 4\% and 3\% respectively. However, in the days that followed the election, the market quickly bounced back \cite{kuriloff_driebusch_2016, points_brushed}. November 9, 2016 08:00:00 GMT was used as the event time for analysis as this was the approximate time at which various news networks announced that Trump had won the election. 

\begin{figure*}
\includegraphics[width = 0.95\textwidth]{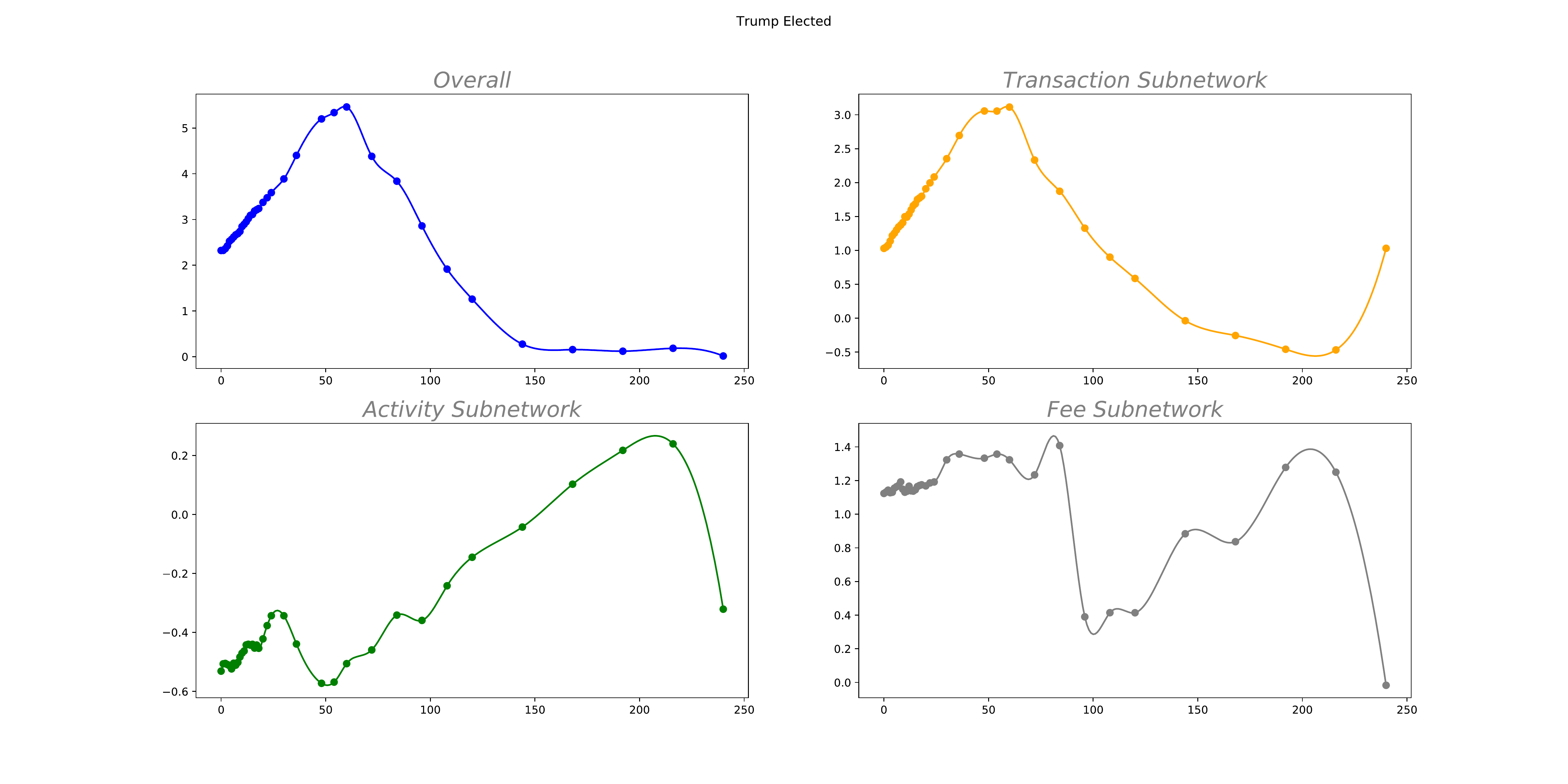}
\caption{Network and subnetwork changes associated with the 2016 Election of Trump. The x axis depicts the hours after the event and the y axis depicts the I-Score value.}
\label{fig:electionTemporal}   
\end{figure*}
As shown in Fig. \ref{fig:electionTemporal}, this event is associated with a substantial change in the overall Bitcoin blockchain network and the transaction subnetwork with a smaller change in the fee subnetwork. However, it did not cause a significant measurable change in the activity subnetwork. While this event is associated with an immediate change in overall blockchain and transaction networks, the greatest measurable change (as indicated by the I-Score) of these two networks was 60 hours after the event (comparing the blockchain network 60 hours before the event with the blockchain network 60 hours after the event). We believe that this gradual, delayed network change is due to the somewhat-continuous nature of this event. Furthermore, we believe that the high I-score of this event is the result of the relatively event-free background time period in late 2016 (when compared to the eventful background time period of other events in late 2017). 

\subsubsection{Overall Bitcoin Blockchain Network Impact}
\label{sec:electionoverall}
\begin{figure}
\includegraphics[width = 0.45\textwidth]{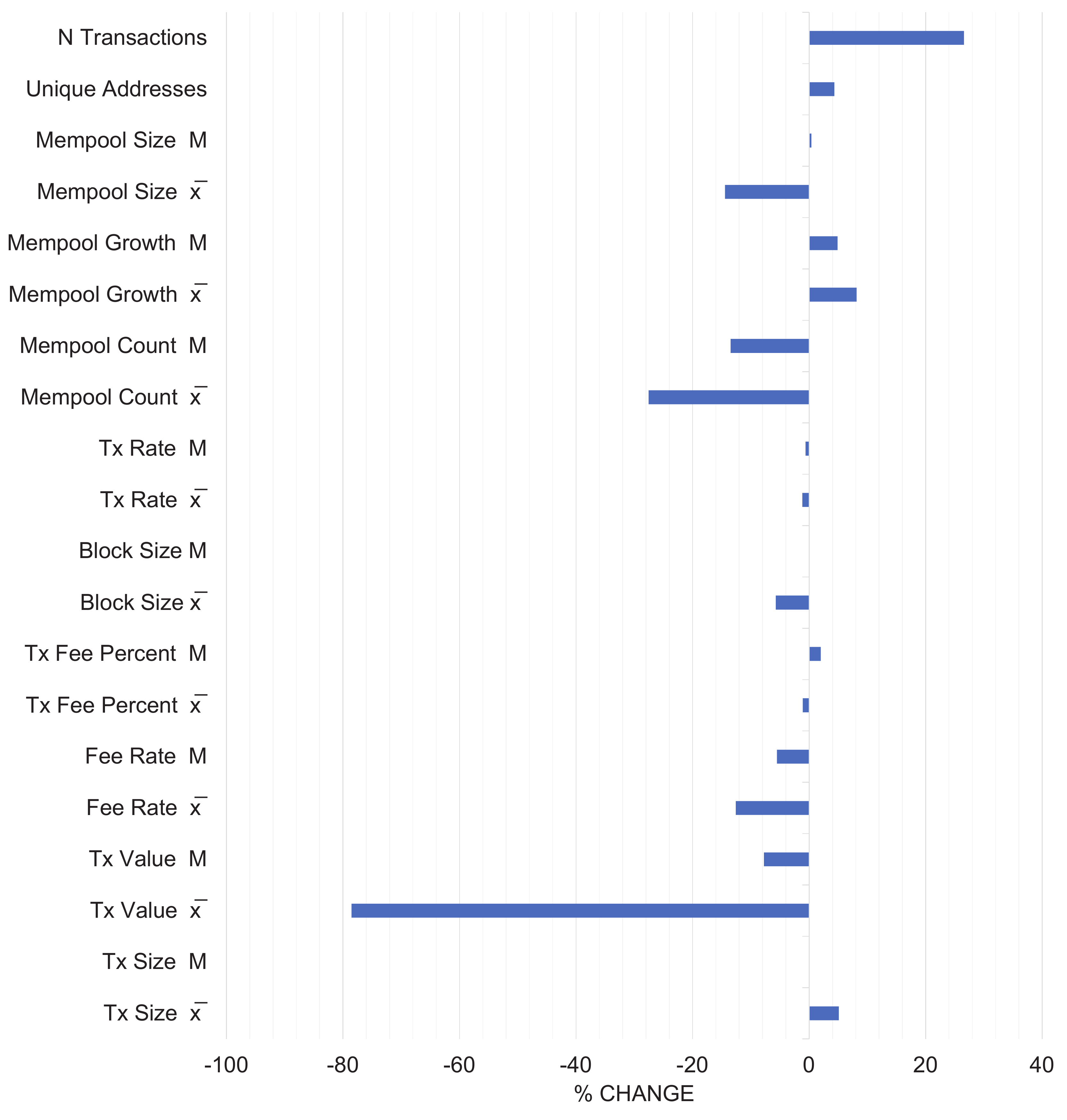}
\caption{Overall Bitcoin Blockchain Network 54 hours after the 2016 Election of Trump versus 54 hours before}
\label{fig:electionFchange}   
\end{figure}

As shown in Fig. \ref{fig:electionFchange}, a notable change associated with the Election of Trump is a significant decrease in the average transaction value. This drop was due to a sharp decrease in the number of high-value transactions after the election. Other notable changes include a 13.47\% drop in median mempool count, which indicates that there were more transactions waiting to be confirmed before the election result, and a 26.57\% increase in the total number of confirmed transactions. We believe that the decrease in mempool count coupled with an increase in the total number of confirmed transactions and a similar transaction rate is the result of event buildup and general hype in the days before the election of Trump as well as increased speculation and activity due to the increase in Bitcoin price following the event. 

Interestingly, while there was a 26.57\% increase in the total number of confirmed transactions (N Transactions), the number of unique address only increased 4.33\%. This indicates an increase in activity per entity which may be a result of an increase in existing entity activity as opposed to new entity activity.

\subsubsection{Transaction Subnetwork Impact}
\label{sec:electiontx}
\begin{figure}
\includegraphics[width = 0.45\textwidth]{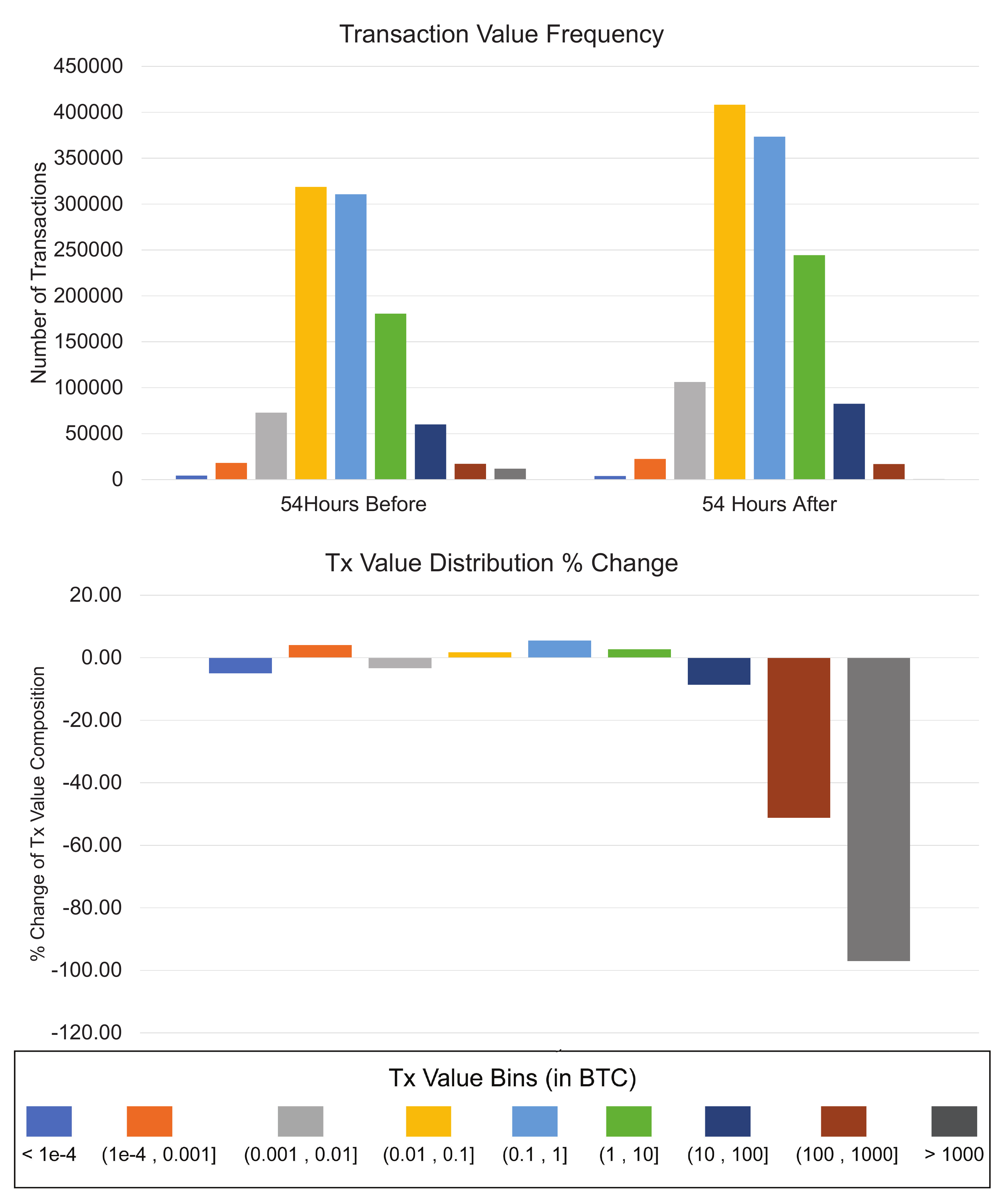}
\caption{Election of Trump Bitcoin Transaction Value Distributions}
\label{fig:electionTx}   
\end{figure}

Analysis of the transaction subnetwork 54 hours before and after the Election of Trump revealed significant changes in the distribution of the values of transactions (in Bitcoin) as shown in Fig. \ref{fig:electionTx}. At 54 hours after the Election of Trump, there was a 51.21\% decrease in the proportion of transactions with values (10, 100] BTC $\approx$ \$(7000, 70000] USD and a 97.07\% decrease in the proportion of transactions with values greater than 1000 BTC $\approx$ \$700,000 USD. Despite the 26.57\% increase in number of confirmed transactions, the number of transactions with values greater than 1000 BTC dropped from 11,893 transactions before to a mere 545 transactions after. 

\subsection{Bitcoin-Bitcoin Cash (BTC-BCH) Hard Fork}
\label{sec:fork}
The Bitcoin-Bitcoin Cash fork was the largest planned hard fork in the Bitcoin history (and blockchain technology in general). It created a new variant of Bitcoin called Bitcoin Cash (BCH) which exists as a new currency in addition to Bitcoin (BTC) (note that Bitcoin and Bitcoin Cash cannot be used interchangeably). The fork arose as a result of disagreement of the block size cap where proponents of Bitcoin Cash believed in a larger block size cap to alleviate scaling issues as Bitcoin became more popular. Ultimately the issue stemmed from high transaction fees as the block size cap prevented more transactions from added to the blockchain which lead to a greater number of transactions in the bitcoin mempool and hence higher transaction fees. As a result, the entire Bitcoin blockchain was cloned and certain rules were changed on August 1, 2017 to create Bitcoin Cash (and its respective blockchain) with a max block size of 8 MB versus the 1 MB of Bitcoin (at that time). The Bitcoin blockchain split into two different blockchains: one Bitcoin and the other Bitcoin Cash, each with its own market, mining network, etc. After the fork, owners of Bitcoin now held an equivalent amount of Bitcoin Cash in addition to their Bitcoin. Due to the planned nature of this event, the exact time of the event was known to be August 1, 2017 12:37 GMT (Block 478,558). August 1, 2017 12:00:00 GMT was used as the event time for analysis. 

\begin{figure*}
\includegraphics[width = 0.95\textwidth]{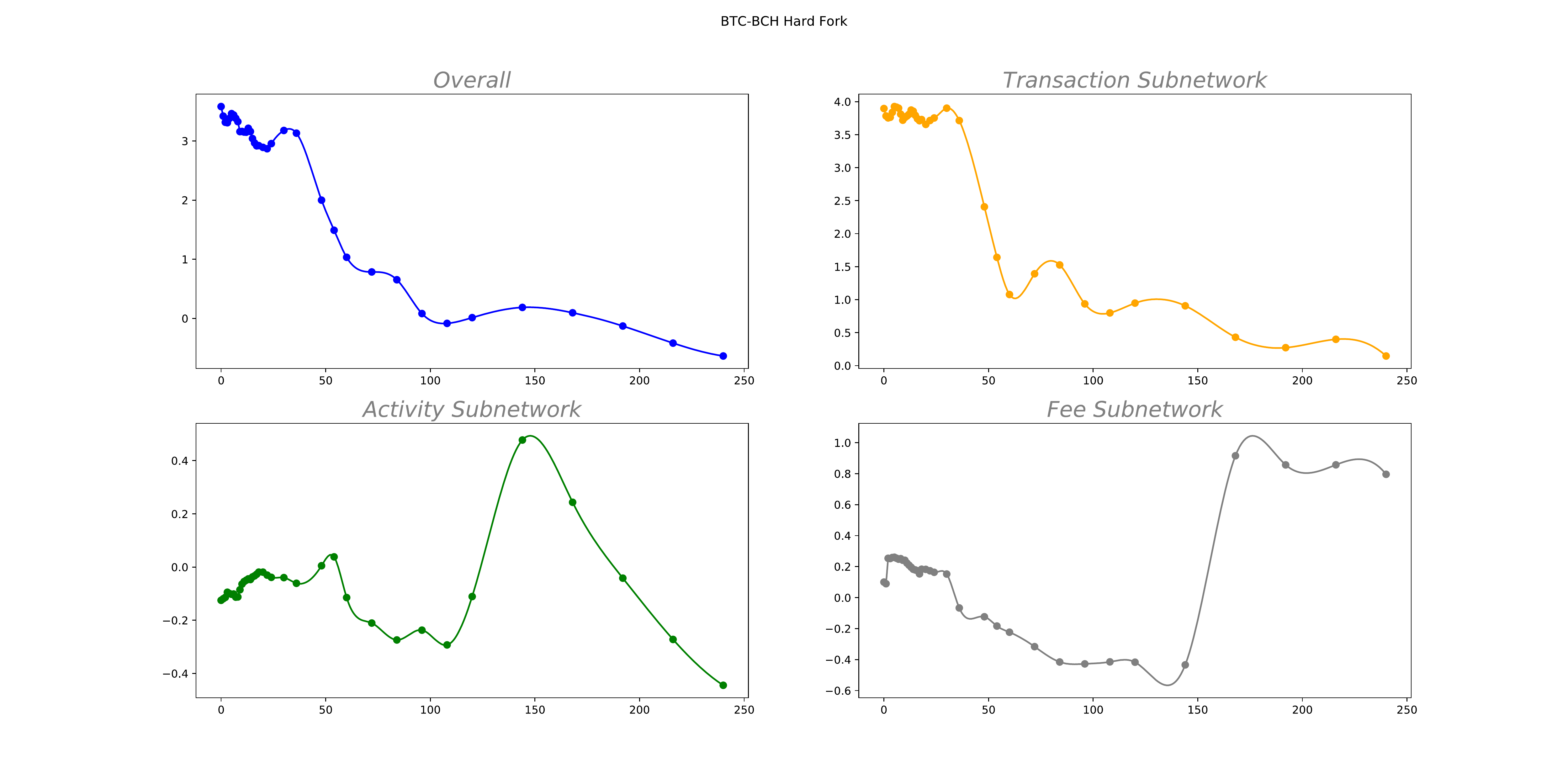}
\caption{Network and subnetwork changes associated with the 2016 Election of Trump. The x axis depicts the hours after the event and the y axis depicts the I-Score value.}
\label{fig:forkTemporal}   
\end{figure*}

Like the election of Trump, the BTC-BCH hard fork caused a significant change in the overall Bitcoin blockchain network and the transaction subnetwork as shown in Figure \ref{fig:forkTemporal}. However, it did not cause a significant measurable change in the fee or activity subnetworks. However, in contrast to the election of Trump, this event caused an immediate change to the overall network and the transaction sub network with the greatest change (highest I-Score) observed for the overall network 0 hours after the event. We believe that due to the discrete nature of the event (it had a well-defined, pre-planned time) there was an immediate reaction and change in the overall Bitcoin blockchain network and the transaction subnetwork.  

\subsubsection{Overall Bitcoin Blockchain Network Impact}
\label{sec:forkOverall}
\begin{figure}
\includegraphics[width = 0.5\textwidth]{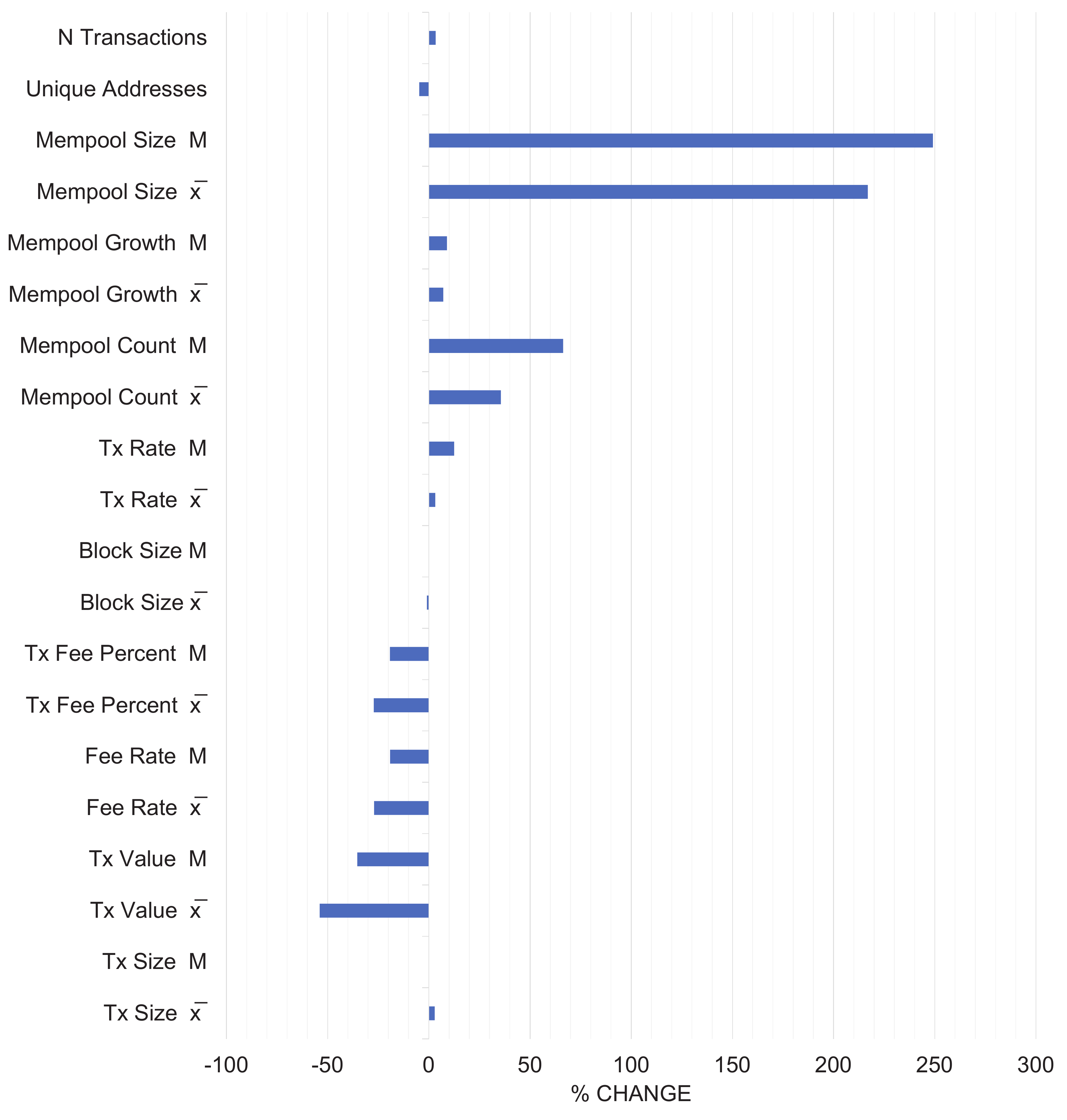}
\caption{Overall Blockchain Network 0 hours after the BTC-BCH Hard Fork versus 0 hours before}
\label{fig:forkFchange}   
\end{figure}

As shown in Figure \ref{fig:forkFchange}, there was a 53.94\% decrease in the values of transactions as well as a 19.81\% decrease in median transaction fees directly after the hard fork. Furthermore, there was a 249.07\% increase in median mempool size and a 66.38\% increase in mempool count (number of unconfirmed transactions) after the hard fork. We believe that the dramatic increase in Mempool Size and Count was the direct result of a massive flood of transaction activity as people sold BTC or BCH to buy the other coin which they believed would succeed (BTC supporters sold their BCH to buy BTC, BCH supporters sold their BTC to buy BCH).  Note that the reason there was not a substantial increase in the number of confirmed transactions (N Transactions) was because the 1 MB block size cap in Bitcoin (at that time) which capped the number of confirmed transactions in a given period of time (median block size before and after the event was around 0.998 MB). Furthermore, the slight decrease in number of unique addresses is likely due to the implementation of our analysis framework, which is only able to count addresses of confirmed transactions (our implementation cannot count the addresses from the unconfirmed transactions in the mempool).

\subsubsection{Transaction Subnetwork Impact}
\label{sec:forktransaction}
\begin{figure}
\includegraphics[width = 0.5\textwidth]{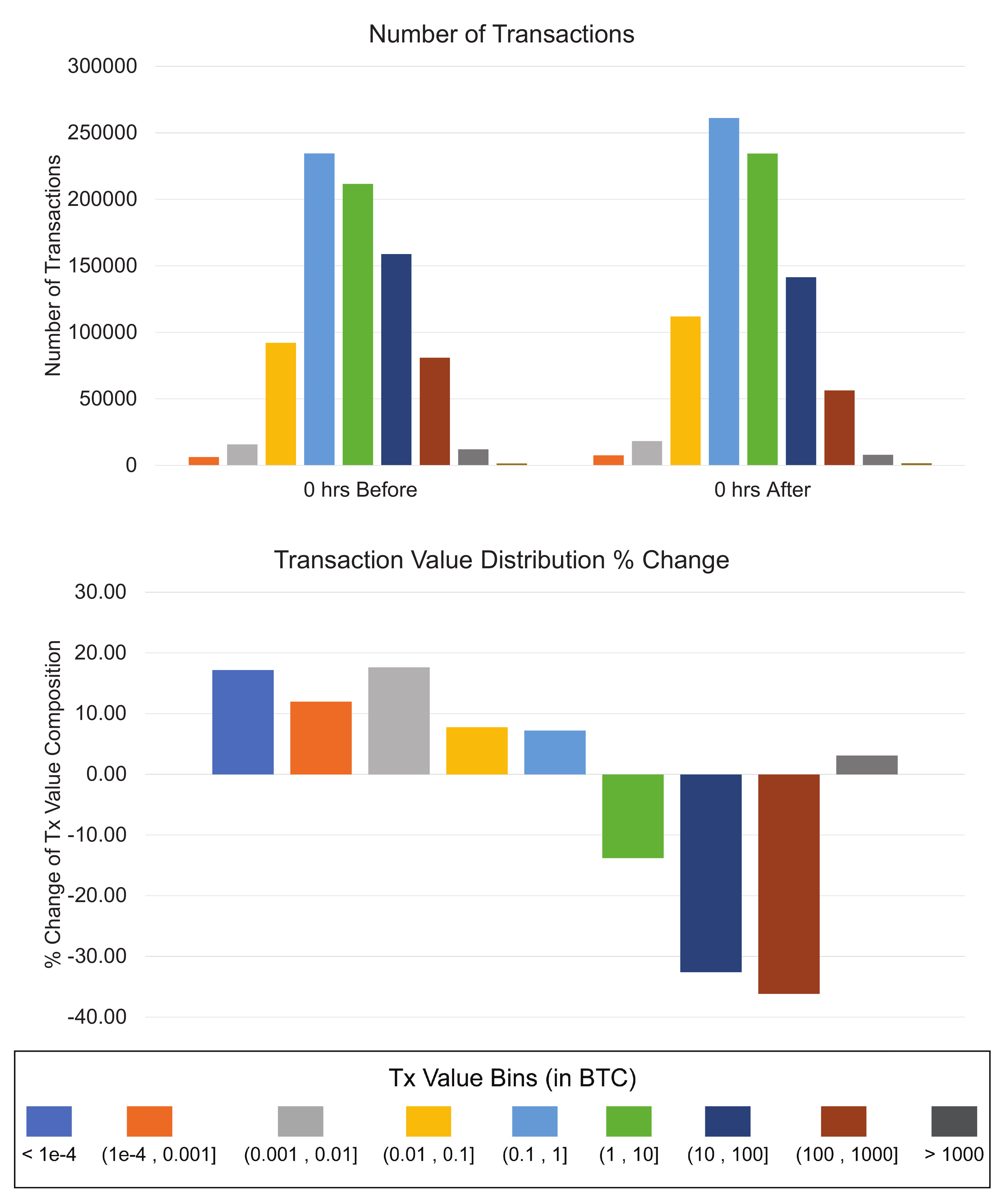}
\caption{BTC-BCH Hard Fork Transaction Value Distributions}
\label{fig:forkTX}   
\end{figure}

The transaction subnetwork showed significant change when comparing the subnetwork directly after the event to before the event. As shown in the distribution of transaction values in Figure \ref{fig:forkTX}, there was a general increase in the number and proportion of transactions with values under 1 BTC $\approx$ \$2,700 USD and a general decrease in number and proportion of transactions with values over 1 BTC (apart from transactions with values greater than 1000 BTC). We postulate that the increase in small-value transactions (values under 1 BTC) is the result of the increase in general transaction activity caused by the fork as people sold or bought BTC to switch to or from BCH. As a corollary, the decrease in transactions with values greater than 1 BTC was likely the result of uncertainty regarding the future success of either cryptocurrency and uncertainty in the BTC and BCH markets.

\subsection{South Korea announces potential Bitcoin ban}
\label{sec:SKBan}
\begin{figure*}
\includegraphics[width = 0.95\textwidth]{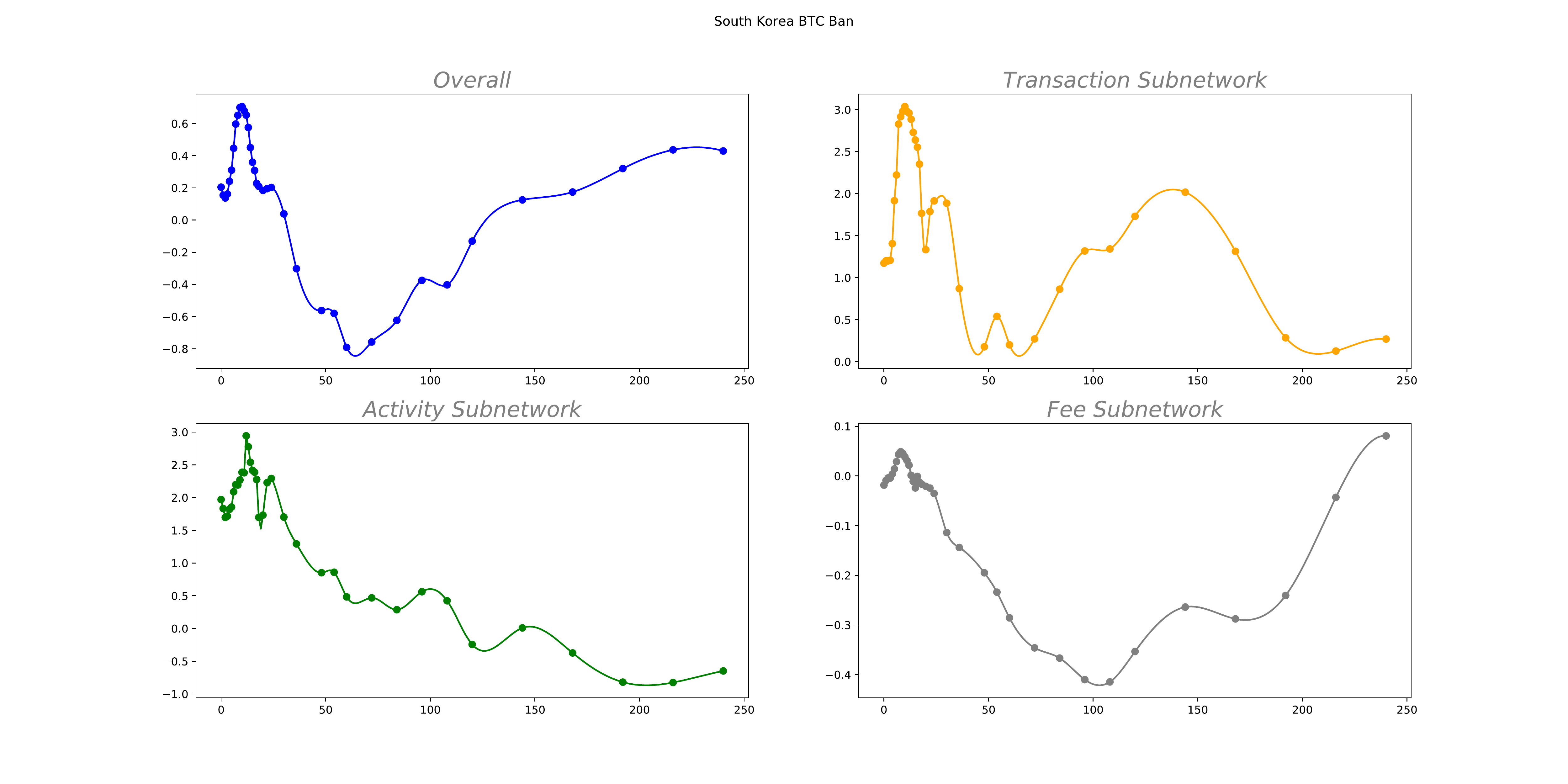}
\caption{Network and subnetwork changes associated with the announcement of the ban }
\label{fig:SKtemporal}   
\end{figure*}

Bitcoin exploded in popularity in South Korea in 2017 with public interest and general hype exceeding that of other nations. On January 11, 2018, justice minister Park Sang-ki of South Korea announced a possible ban on cryptocurrency trading in South Korea. As a result of this announcement, the price of Bitcoin fell \$2000 in value from ~\$15,000 USD and the Bitcoin-Won exchange rate dropped ~21\% \cite{zhong_2017}. Jan 11, 2018 12:00:00 GMT was used as the event time for analysis.
As shown in Figure \ref{fig:SKtemporal}, there a substantial change in the transaction and activity subnetworks. There was also a substantial change in the overall network however the I-Score metric does not reflect this because the time period (November 2017 February 2018) used to calculate the background fluctuations was a period of intense Bitcoin speculation and frenzy (leading to the price peak in December of over \$19,000 for a Bitcoin) with hundreds of Bitcoin events. As noted earlier, the I-Score metric underestimates event-induced blockchain network change during eventful time periods. Similar to the election of Trump, there was a delayed maximum network change approximately 12 hours after the announcement of the ban which we believe is the result of the timing of news reports and the inherent delay as people understand/react to the event.
We postulate that the second spike in I-Score as shown in  the overall and transaction network approximately 5 days following the initial announcement of the ban (shown in Figure \ref{fig:SKtemporal}) is the result of further news regarding cryptocurrency regulation in South Korea (including a new requirement for Koreans to use their real names on cryptocurrency accounts which defeats the anonymity of Bitcoin) and a clarification that the announced cryptocurrency ban was far from a final decision.

\subsubsection{Overall Bitcoin Blockchain Network Impact}
\label{sec:SKoverall}
\begin{figure}
\includegraphics[width = 0.45\textwidth]{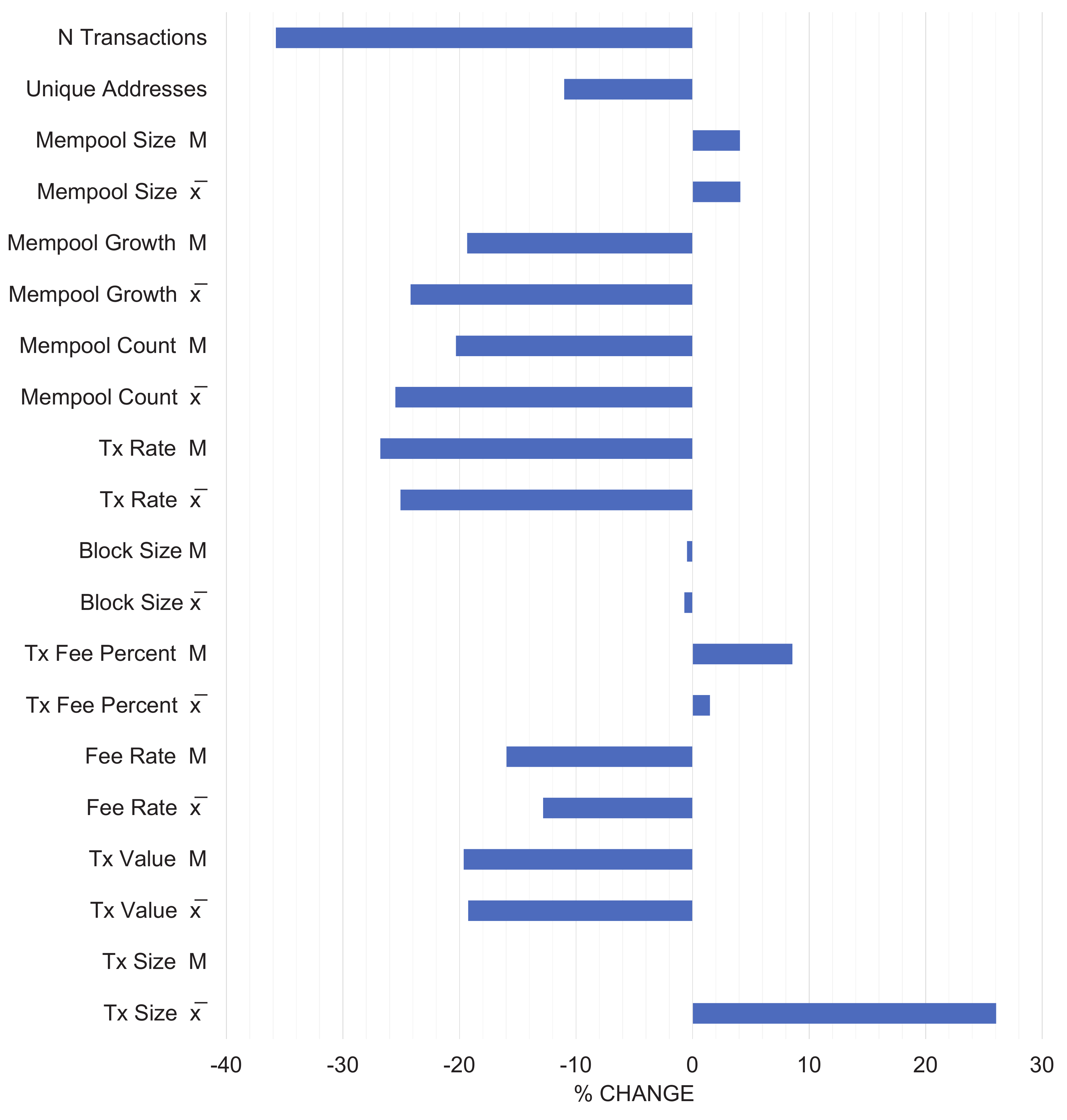}
\caption{Overall blockchain network 12 hours after the announcement of the ban versus 12 hours before}
\label{fig:SKFchange}   
\end{figure}

Analysis 12 hours after the announcement of the ban reveals a substantial decrease overall activity as shown in Figure \ref{fig:SKFchange}(note that the network state analyzed here is after the substantial Bitcoin price drop that immediately followed the event).  Overall, there is a drastic decrease in activity 12 hours after the event as shown by the 26.8\% decrease in median transaction rate, a 20.31\% decrease in median mempool count, a 11.02\% decrease in the number of unique addresses and a 35.75\% decrease in confirmed transactions from ~1.42 million before to ~910,000 after the announcement of the ban. We believe a key contributing factor to this decrease in activity was the bleak outlook for Bitcoin trading coupled with the substantial price drop as a result of the fear, uncertainty, and doubt (FUD) caused by the event.

\subsubsection{Transaction Subnetwork Impact}
\label{sec:SKtx}
\begin{figure}
\includegraphics[width = 0.45\textwidth]{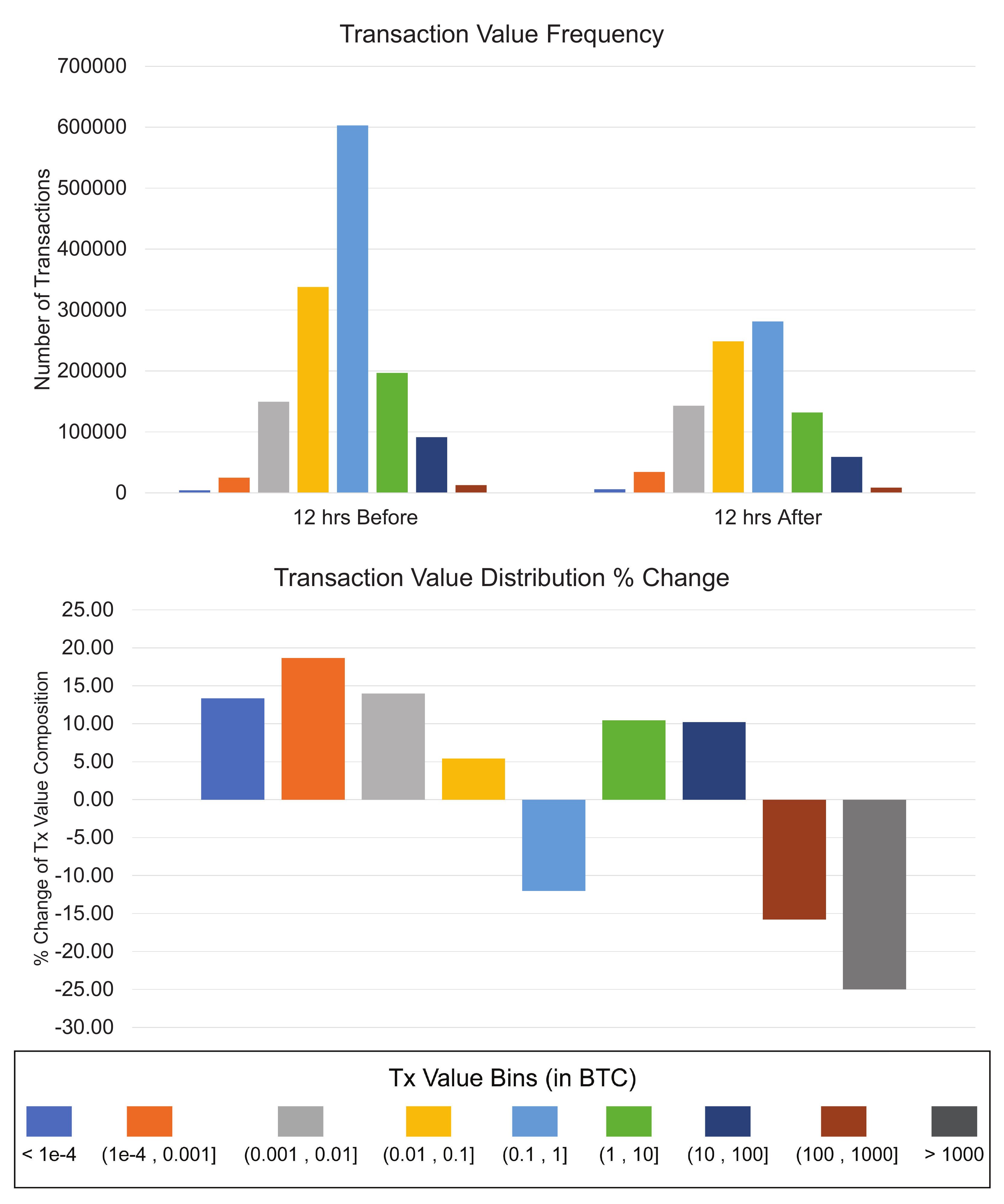}
\caption{South Korea BTC Ban Announcement Transaction Value Distributions}
\label{fig:SKtxchanges}   
\end{figure}

With regards to transaction values, there was a change in the distribution of values after the announcement as shown in Figure \ref{fig:SKtxchanges}. There was an approximately 15-25\% decrease in the proportion of transactions with value greater than 100 BTC $\approx$ \$130,000 USD and a approximately 10-20\% increase in the proportion of transactions with values smaller than 0.01 BTC $\approx$ \$130 USD. We postulate that the FUD caused by this announcement are the key factors behind the decrease in the proportion of high-value transactions much like the decrease in high-value transactions after the Election of Trump. Interestingly, while the overall number of transactions decreased substantially after the event, the number of transactions with values less than 0.001 BTC $\approx$ \$13 BTC actually increased by approximately 38\%. We believe this can be partly explained by regular gambling or every day BTC to BTC transactions which are rather unaffected by the outlook or uncertainty of Bitcoin in South Korea. 

\section{Discussion}
\label{sec:discussion}
The 16 events between 2016-2018 selected for analysis is not a complete set of all events during this time period. Thus, if our process is robust, we should observe spikes in distance from 2016-2018 associated with events outside our selected set. Furthermore, a large majority of spikes in distance should be able to be associated with events. 
To establish the robustness of our process, we examine 2016-2018 with rolling blockchain network structure distance analysis from May 2016 - June 2018 using a constant data frame length of 96 hours and a gap length of 0 hours. Note that this process biases towards events with immediate significant change such as the BTC-BCH Hard Fork and may leave out some events with a more gradual change (i.e. Bitcoin price peak). 
\subsection{Overall Bitcoin Blockchain Network}
\label{sec:Overalldiscussion}
\begin{figure*}
\includegraphics[width = 0.98\textwidth]{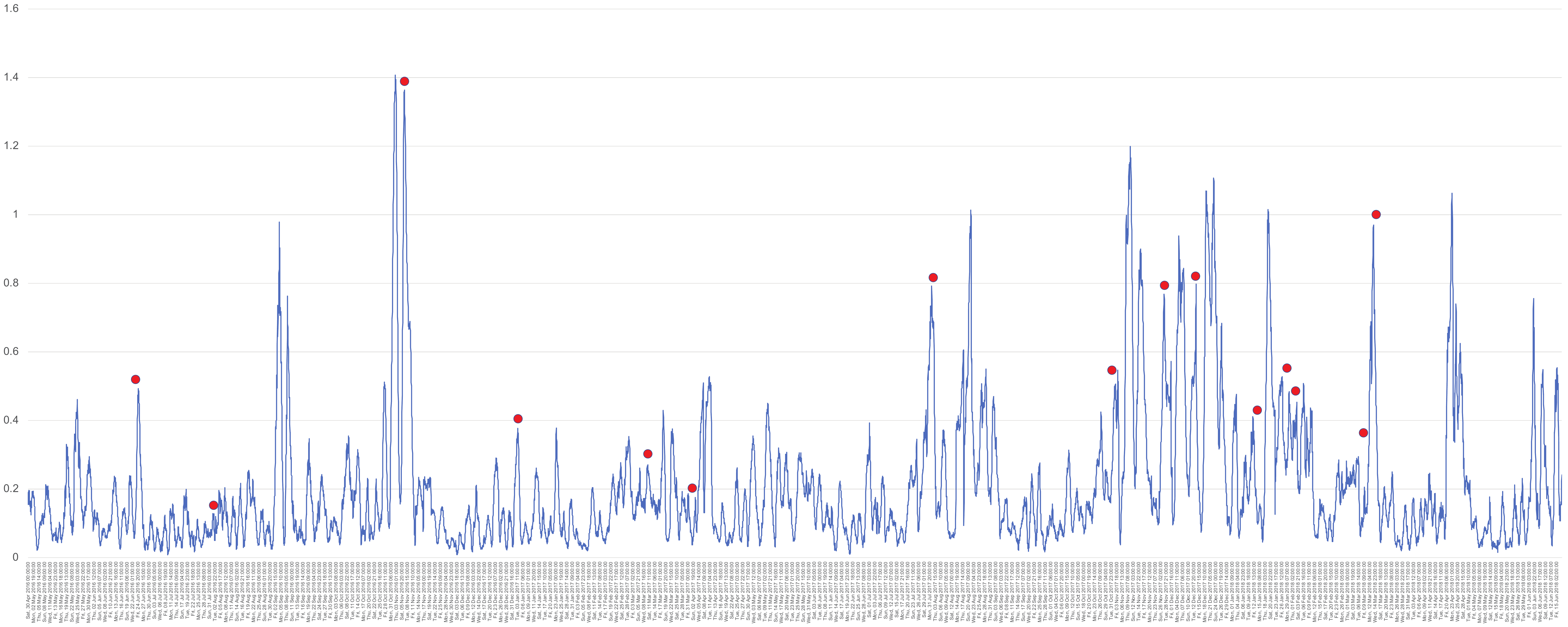}
\caption{Change in the Overall Bitcoin Blockchain Network from 2016-2018. A larger distance value (shown in the y axis) indicates greater network change at that time.}
\label{fig:OverallRetrospective}   
\end{figure*}

As shown in Figure \ref{fig:OverallRetrospective}, there are many spikes in the overall blockchain network structure distance that were not associated with the events selected which are indicated by the circles (note that this overall analysis does not use the I-score, rather it uses the distance measure value directly). However, upon further investigation, we were able to find events using Google News that correlate to a majority of the discernible spikes with distance greater than 0.4 as shown in Figure \ref{fig:OverallRetrospective}. 

\begin{figure*}
\includegraphics[width = 0.95\textwidth]{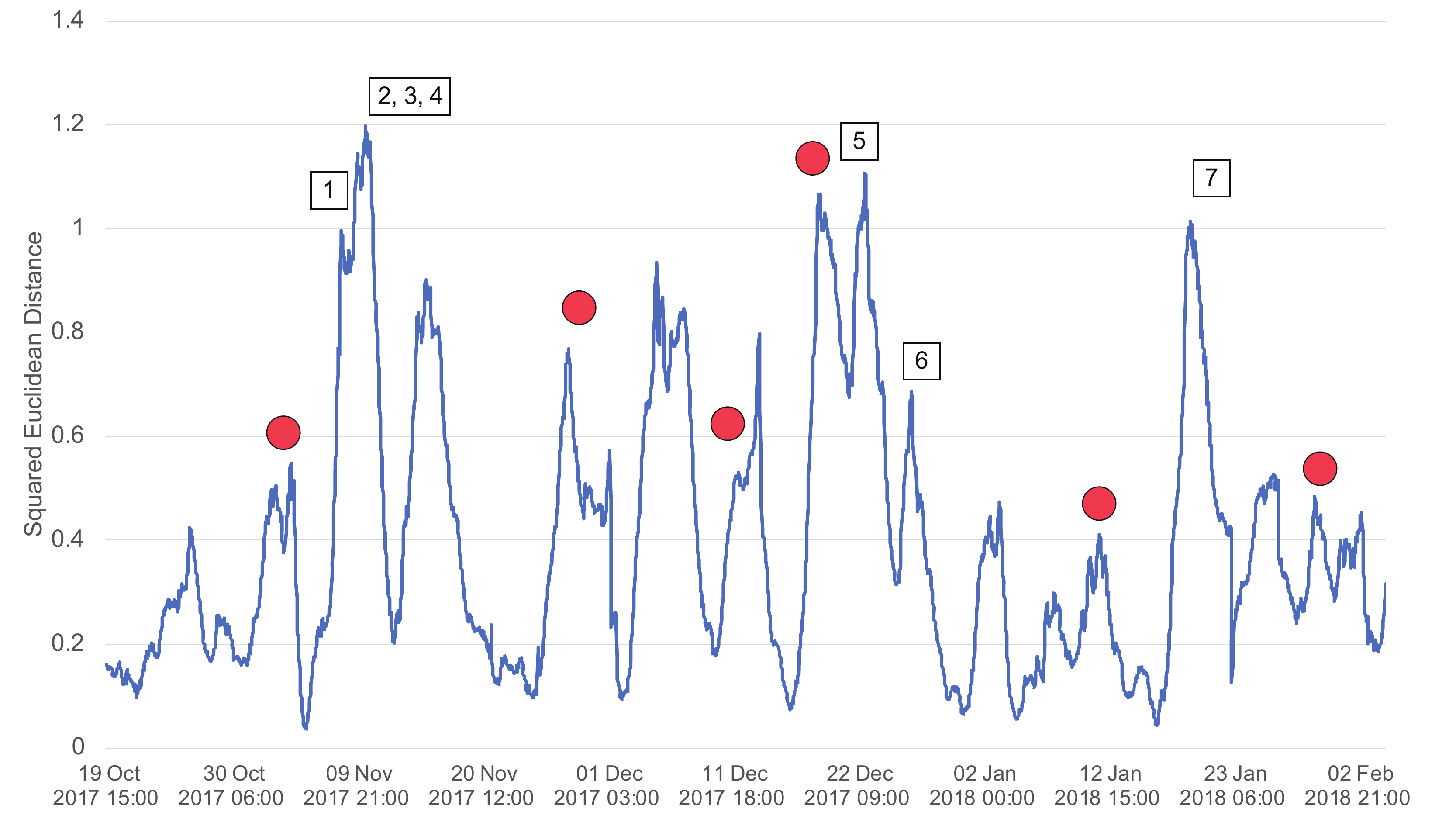}
\caption{Overall Bitcoin Blockchain network change from Oct 2017-Feb 2018}
\label{fig:OverallActivechange}   
\end{figure*}

In particular, we analyze the extremely eventful time period from October 2017- February 2018 to see if most observed spikes in distance can be correlated to events. As shown in Figure \ref{fig:OverallActivechange} and Table \ref{tab:activeevents}, we are able to correlate events to a vast majority of spikes in distance observed. Note that some spikes are associated with multiple events as our process is unable to conclusively associate impact of a single event out of a set of significant events that occur within a short time span. red dot indicates that the event was part of our case study of 16 events. 

\begin{table*}
\caption{Overall Bitcoin Blockchain Network Distance Associated Events}
\label{tab:activeevents}
\centering
\arrayrulecolor{black}
\begin{tabular}{!{\color{black}\vrule}S|W!{\color{black}\vrule}Z!{\color{black}\vrule}} 
\hline
\textbf{ N}   & \textbf{ Time}  (Approx. Range)  & \textbf{ Associated Event ~ }                                                                                        \\ 
\hline
1             & Nov 8, 2017                      & Segwit2X Fork is cancelled due to lack of consensus \cite{hertig_hertig_2017}.                                                                 \\ 
\arrayrulecolor{black}\hline
2             & Nov 11-18, 2017                  & The rivalry between Bitcoin Cash and Bitcoin continues as BCH/BTC prices fluctuate wildly.                            \\ 
\hline
3             & Nov 11-18, 2017                  & Bitcoin Gold, another Bitcoin fork, is launched on Nov 12.                                                             \\ 
\hline
4             & Nov 17, 2017                     & Bitcoin price breaks \$8000.                                                                                          \\ 
\hline
5             & Dec 22, 2017                     & Bitcoin prices drop over 30\% in one day \cite{popper_hsu_2017}.                                                                             \\ 
\hline
6             & Dec 28, 2017                     & South Korea proposes Bitcoin trading regulation \cite{lee_zhong_2018}.                                                                      \\ 
\hline
7             & Jan 17-25, 2018                  & Bitcoin price falls below \$10,000 in response to regulation and cases of fraud in the US, China, South Korea, etc. \cite{popper_bowles_2018_bubbledeflate}  \\
\arrayrulecolor{black}\hline
\end{tabular}
\end{table*}

\begin{figure*}
\includegraphics[width = 0.95\textwidth]{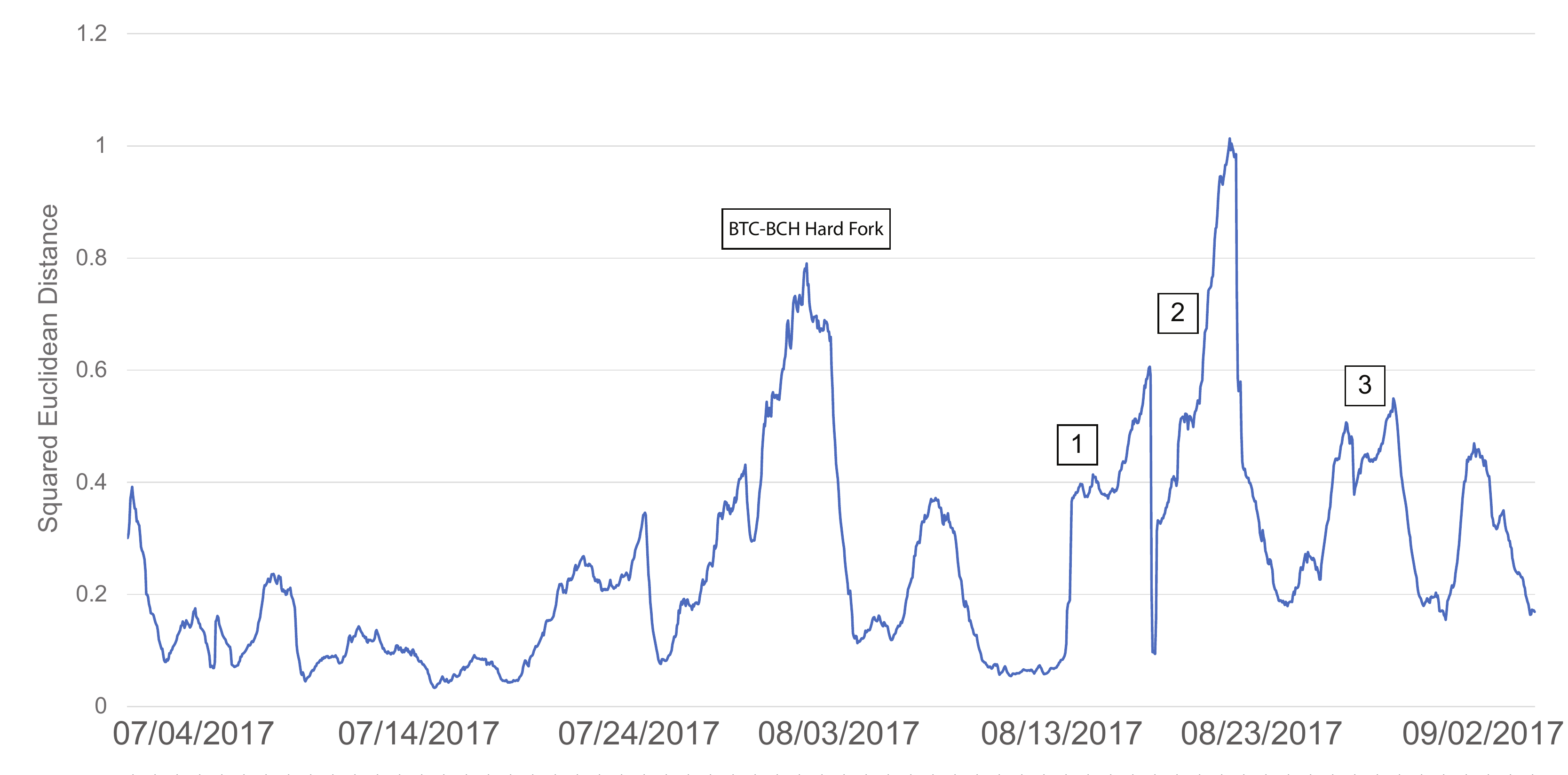}
\caption{Overall Bitcoin Blockchain Network Change before and after the BTC-BCH fork}
\label{fig:hardforkretrospective}   
\end{figure*}

While we cannot conclusively associate these events with the spikes in distance, the events we associated in Table \ref{tab:activeevents}.  were the most significant/received the most media attention around the time of the spikes. 
From Figure \ref{fig:OverallActivechange}, we also observe latent ‘aftershocks’ associated with an event. In particular, in the month after the BTC-BCH hard fork, we observe several latent spikes associated with various hard fork related news or developments. As shown in Figure \ref{fig:hardforkretrospective}, the hard fork on August 1 was preceded by a buildup in network change and followed by three distinct ‘after-shocks’. 

\textit{Box 1:} On August 14, Bitcoin reached a new record high of over \$4300 USD = 1 BTC in the wake of the fork prompting a flurry of news and speculation \cite{popper_2017_pricesurge)}. \\
\textit{Box 2:} Bitcoin miners swapped back and forth between mining Bitcoin and Bitcoin Cash as difficulty and profitability fluctuates \cite{song_song_2017}. \\
\textit{Box 3:} Bitcoin reaches a record high price (increasing profitability) while Bitcoin Cash (the forked currency) reaches a “11-day low” \cite{chengevelyn_2017_jump70}

\subsection{Fee Subnetwork}
\label{sec:FeeRetrospective}
\begin{figure*}
\includegraphics[width = 0.95\textwidth]{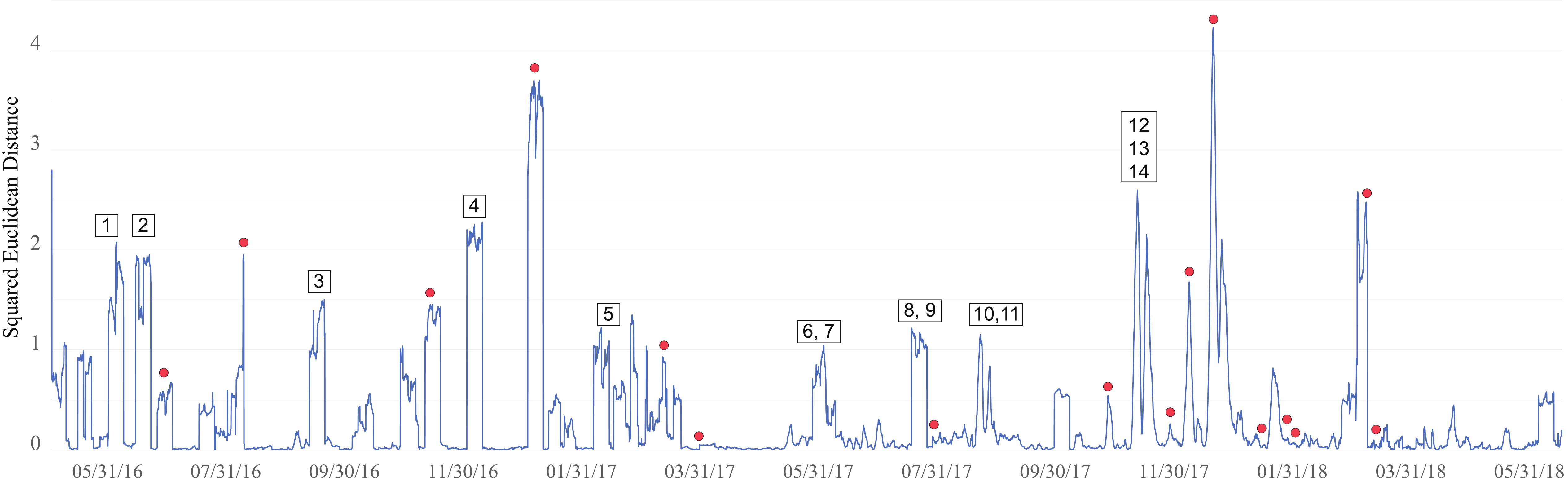}
\caption{Fee Subnetwork change from 2016-2018}
\label{fig:feeretrospective}   
\end{figure*}

In the fee subnetwork, we are able to correlate events to almost every major spike in fee subnetwork change as shown in Figure \ref{fig:feeretrospective} and Table \ref{tab:feeretroevents}. The only exception is the change in the fee subnetwork (Box 3) from Sep. 10-15, 2016. 

\begin{table*}
\centering
\caption{Fee Subnetwork Distance Associated Events}
\label{tab:feeretroevents}
\arrayrulecolor{black}
\begin{tabular}{!{\color{black}\vrule}S!{\color{black}\vrule}W!{\color{black}\vrule}Z!{\color{black}\vrule}} 
\hline
\textbf{ N  } & \textbf{ Time  }(Approx. Range)\textbf{ } & \textbf{ Associated Event  }                                                                                                                         \\ 
\hline
1             & May 29, 2016                              & BTC price breaks \$500                                                                                                                               \\ 
\hline
2             & Jun 17, 2016                              & A \$50 million-dollar hack occurs on Ethereum, a “rival” cryptocurrency \cite{popper_2016_daohack}.                                                                              \\ 
\hline
3             & Sep 10-15, 2016                           &  \textit{Unknown}                                                                                                                                    \\ 
\hline
4             & Nov 29-Dec 7, 2016                        &  \textit{Unclear}: 1) IRS audits Coinbase, a major cryptocurrency exchange \cite{roberts_2016_IRS}. 2) Bitcoin is predicted to rise as a result of the election of Donald Trump \cite{shen_2016}.    \\ 
\hline
5             & Feb 9, 2017                               & Two major Chinese Bitcoin exchanges stop withdrawals \cite{deng_2017}.                                                                                                 \\ 
\hline
6             & May 20-28, 2017                           & Bitcoin price breaks \$2000 for the first time, peaking at \$2700 before falling 30\% in a day.                                                       \\ 
\hline
7             & May 24, 2017                              & CEO of Fidelity announces support for Bitcoin \cite{krouse_2017}.                                                                                                        \\ 
\hline
8             & July 11-17, 2017                          & Cryptocurrency prices collapse (Bitcoin loses \textasciitilde{}40\% of its value from all time high)                                                 \\ 
\hline
9             & July 27, 2017                             & BTC-E, a Bitcoin exchange known for money laundering for criminals, is shut down and its ‘mastermind’ is arrested \cite{popper_2017_nexuscrime}.                                    \\ 
\hline
10            & Aug 18-25, 2017                           & Aftershocks from the BTC and BCH hard fork – miners swap back and forth as profitability of each coin fluctuates \cite{song_song_2017}.                                     \\ 
\hline
11            & Aug 29, 2017                              & Bitcoin price reaches new peak of over \$4700 possibly in response to North Korea nuclear threat \cite{chengevelyn_2017_jump70}.                                                     \\ 
\hline
12            & Nov 11-18, 2017                           & The rivalry between Bitcoin Cash and Bitcoin continues as BCH/BTC prices fluctuate wildly                                                            \\ 
\hline
13            & Nov 11-18, 2017                           & Bitcoin Gold, another Bitcoin fork, is launched on Nov 12.                                                                                             \\ 
\hline
14            & Nov 17, 2017                              & Bitcoin price breaks \$8000.                                                                                                                          \\
\hline
\end{tabular}
\arrayrulecolor{black}
\end{table*}

We cannot conclusively link these events with the spikes in distance or draw a conclusive cause-effect relationship. However, the events we associated in Table \ref{tab:feeretroevents}. were the most significant/received the most media attention around the time of the spikes. 
A major observation we can draw from our retrospective analysis of Bitcoin fee subnetwork changes is that a vast majority of major fee changes are correlated with Bitcoin financial-type events. Furthermore, we observe that Bitcoin price-milestones (breaking \$1000, \$2000, etc.) are associated with a change in the Bitcoin fee subnetwork. This is consistent with expectations as a price milestone will generally increase demand for Bitcoin thus driving up fees as more transactions are broadcast (as people buy/sell at this milestone). 

\subsection{Discussion on I-Score and Framework}
\label{sec:IScoreDiscussion}
The I-score process presented in this paper represents a first step towards attempting to quantifying changes and shifts in the Bitcoin blockchain network. It is difficult to empirically validate the effectiveness of the I-score and our analysis framework due to the lack of a ground truth for comparison in this case study. Furthermore as an observational case study, we cannot directly quantifying and prove event impact. Due to the constantly evolving and dynamic nature of the Bitcoin blockchain and the cryptocurrency, it is also extremely hard to create theoretically-proven methodology of measuring changes and shifts. However, we are able to demonstrate the effectiveness of the I-score framework in several ways. 

As demonstrated in the Section \ref{sec:Overalldiscussion}  and \ref{sec:FeeRetrospective}, we applied our I-score process in a retrospective analysis and were able correlate major events with a majority of spikes observed in I-score from 2016-2018. This minimal false positive rate is one aspect that lends credibility to the I-score process. Furthermore, many of the spikes in I-score in the retrospective analysis were consistent with logical expectation (e.g. the I-score spikes in fee subnetworks associated with major Bitcoin price-milestones). 

The results derived from the I-score process presented in Section \ref{sec:casestudy} are also largely consistent with logical expectation. Notably, the remarkable similarity between the two temporal I-score graphs observed in the two exchange price milestone events (shown in Fig. \ref{fig:btcmilestone}) demonstrates consistent results produced by the I-score across extremely similar events. 

\section{Conclusions}
\label{sec:conclusion}
The contribution of this paper is twofold. First, we introduce a general framework to quantitatively analyze holistic and specific changes which can be applied to many cryptocurrencies. Our framework is computationally-efficient for tackling the two challenges: capturing the network as a whole and separating event changes from natural fluctuations. The I-Score metric facilitates distinguishing event-induced changes from natural/non-event fluctuations in blockchain ecosystems. It allows for cross-comparisons across all configurations (networks or subnetworks, time intervals, etc.) of a given blockchain or cryptocurrency ecosystem. We believe that the I-Score metric can be used to evaluate impacts of events on many other cryptocurrencies. Second, we have applied our framework to the Bitcoin ecosystem to demonstrate the impact of certain classes of events on particular aspects of the Bitcoin blockchain network. We observed roughly generalizable correlations between specific event types and shifts in Bitcoin subnetworks. Events within three subgroups have strongly consistent impacts on blockchain network/subnetworks. Extended analysis of three events has revealed specific changes in Bitcoin transaction value distributions and other Bitcoin blockchain network indicators. For robustness, we demonstrate that a majority of substantial changes observed using our framework can be associated with events. Retrospective analysis also reveals further correlations between certain event types (such as Bitcoin price milestones) and subnetwork impact. We hope that our research will inform developers and regulators in the development and regulation of future blockchain technologies and the blockchain industry as a whole. 

Our future work will focus on improving the I-score process or replacing it with a more robust process. The I-Score metric assumes a relatively event-free background time period, and thus may have underestimated event-induced changes. For example, the I-Scores for events from late-2017 – early 2018 are likely underestimated because this is an extremely eventful time period. Further improvements to the I-score include an improved statistical method for separation of event-induced changes from natural fluctuations and the addition of other Bitcoin ecosystem indicators to our framework. Unsupervised machine learning will also be explored to analyze event impact on the Bitcoin blockchain network. Specifically, clustering algorithms such as Gaussian Mixture Models or anomaly detection algorithms such as Random Isolation Forest could be explored to tackle this problem. 

\begin{acknowledgements}
We would like to acknowledge high-performance computing support of the R2 compute cluster (DOI: 10.18122/B2S41H) provided by Boise State University’s Research Computing Department.
\end{acknowledgements}

\section*{Conflict of interest}
The authors declare that they have no conflict of interest.

\bibliographystyle{spbasic}
\interlinepenalty=10000
\bibliography{refs}   
\end{document}